\documentclass[journal,transmag]{IEEEtran}
\usepackage{cite}
\usepackage{graphicx}
\usepackage{float}
\usepackage{url}
\usepackage{amsmath,amsfonts,amssymb}
\usepackage{booktabs}
\usepackage{subcaption}
\usepackage[export]{adjustbox}
\usepackage[update,prepend]{epstopdf}
\usepackage[lined,algonl,ruled]{algorithm2e}
\usepackage{balance}
\usepackage{acronym}
\usepackage{color}
\usepackage{multirow}
\usepackage{multicol}
\usepackage{hyperref}
\hypersetup{
	bookmarksopen=true,
	bookmarksnumbered=true,
	colorlinks=true,
	citecolor=blue,
	linkcolor=red,
	linktocpage=true,
	linkbordercolor={1 1 1},
	hypertexnames=false,
	plainpages=true,
	pdfsubject={},
	pdfkeywords={},
	pdftitle={Speech Enhancement  using a Deep Mixture of Experts},
	pdfauthor={Shlomo E. Chazan, Jacob Goldberger and Sharon Gannot}
}

\newcommand{\Log}{\text{Log}}

\newcommand{\xbb}{\boldsymbol{b}}
\newcommand{\rrho}{\boldsymbol{\rho}}
\newcommand{\ttheta}{\boldsymbol{\theta}}

\newcommand{\x}{\mathbf{x}}
\newcommand{\s}{\mathbf{s}}
\newcommand{\n}{\mathbf{m}}

\newcommand{\vv}{\mathbf{v}}

\newcommand{\comment}[1]{}

\ifCLASSINFOpdf

\else

\fi

\acrodef{DSE}{Deep Single Expert}
\acrodef{AIR}{acoustic impulse response}
\acrodef{PSD}{power spectral density}
\acrodef{DNN}{deep neural network}
\acrodef{MFCC}{mel-frequency cepstral coefficients}
\acrodef{MMSE}{minimum mean square error}
\acrodef{ASR}{automatic speech recognition}
\acrodef{STSA}{short-time spectral amplitude estimator}
\acrodef{LSAE}{log-spectral amplitude estimator}
\acrodef{OMLSA}{optimally modified log spectral amplitude}
\acrodef{IMCRA}{improved minima controlled recursive averaging}
\acrodef{STFT}{short-time Fourier transform}
\acrodef{DFT}{discrete Fourier transform}
\acrodef{MoG}{Mixture of Gaussians}
\acrodef{MoE}{Mixture of Experts}
\acrodef{ATF}{acoustic transfer function}
\acrodef{MODE}{mixture of deep experts}
\acrodef{DMoE}{Deep Mixture of Experts}
\acrodef{r.v.}{random variable}
\acrodef{p.d.f.}{probability density function}
\acrodef{NN}{neural network}
\acrodef{EM}{expectation-maximization}
\acrodef{SPP}{speech presence probability}
\acrodef{CMVN}{cepstral mean and variance normalization}
\acrodef{NN-MM}{neural network mixture-maximum}
\acrodef{PESQ}{perceptual evaluation of speech quality}
\acrodef{SNR}{signal to noise ratio}
\acrodef{DAE}{Deep auto-encoder}
\acrodef{LLR}{log likelihood ratio}
\acrodef{WSS}{weighted spectral slope}
\acrodef{Covl}{overall quality}
\acrodef{Csig}{speech distortion}
\acrodef{Cbak}{background distortion}
\acrodef{WSJ}{Wall Street Journal}
\acrodef{SVM}{support vector machine}
\acrodef{IBM}{ideal binary mask}
\acrodef{IRM}{ideal ratio mask}
\acrodef{ReLU}{rectified linear unit}
\acrodef{WER}{word error rate}
\acrodef{MM}{MixMax}
\acrodef{MOS}{mean opinion score}
\acrodef{mse}{mean square error}
\acrodef{MSE}{Mean Square Error}
\acrodef{pDNN}{phoneme DNN}
\acrodef{cDNN}{classifier DNN}
\acrodef{gDNN}{gating DNN}
\acrodef{SGD}{stochastic gradient descent} 
\newcommand{\xcomment}[1]{}

\begin{document}

	\title{Speech Enhancement  using a Deep Mixture of Experts }

	\author{\IEEEauthorblockN{Shlomo E. Chazan,
			Jacob Goldberger and
			Sharon Gannot \IEEEmembership{Senior Member, IEEE,}}
		\thanks{Shlomo E. Chazan, Jacob Goldberger and Sharon Gannot are with the Faculty of Engineering, Bar-Ilan University, Ramat-Gan, 5290002, Israel (e-mail: Shlomi.Chazan@biu.ac.il; Jacob.Goldberger@biu.ac.il; Sharon.Gannot@biu.ac.il).}\\%
		\IEEEauthorblockA{Faculty of Engineering, Bar-Ilan University, Ramat-Gan, 5290002, Israel}}

	\maketitle
	%	\IEEEtitleabstractindextext{
	\begin{abstract}
		In this study we present a \ac{DMoE} neural-network architecture for single microphone speech enhancement. By contrast to most speech enhancement algorithms that overlook the speech variability mainly caused by  phoneme structure, our framework comprises a set of \acp{DNN}, each one of which is an `expert' in enhancing a given speech type  corresponding to a phoneme. A gating  \ac{DNN} determines which expert is assigned  to a given speech segment.  A \ac{SPP} is then obtained as a weighted average of the expert \ac{SPP} decisions, with the weights determined by the gating DNN.  A soft spectral attenuation, based on the \ac{SPP}, is then applied to enhance the noisy speech signal.  The experts and the gating components of the  \ac{DMoE} network are  trained jointly.
 As part of the training,  speech clustering into different subsets is performed in an unsupervised manner. Therefore, unlike previous methods,  a phoneme-labeled dataset is not required for the training procedure.  A series of experiments with different noise types verified the applicability of the new algorithm to the task of speech enhancement. The proposed scheme outperforms other schemes that either do not consider  phoneme structure or use a simpler training methodology.
		%}
	\end{abstract}


	\IEEEdisplaynontitleabstractindextext
	
	\IEEEpeerreviewmaketitle

	\section{Introduction}
	
	\IEEEPARstart{T}{here} are many approaches to solve the problem of speech enhancement using a single channel \cite{38}. Although microphone array algorithms are widely used, there are still applications in which only a single microphone is available. However, the solutions suggested for this task are not always sufficient.

	Classical algorithms such as the \ac{OMLSA} estimator and the  \ac{IMCRA} noise estimator approach to robust speech enhancement were developed to even deal  with nonstationary noise environments~\cite{17,32}. Nevertheless, when input with rapid changes in noise statistics is processed, the estimator tends to yield \emph{musical noise} artifacts at the output of the enhancement algorithm.
	
	Over the past few years fully-connected \ac{DNN}-based algorithms have been developed to enhance noisy speech. \ac{DAE} were  trained to find a non-linear filter between noisy input to clean speech \cite{25}. A set of noisy/clean features constituted the database for the training phase. This approach often suffers from speech distortion when an unfamiliar noise is examined.
	
	In response, an \ac{IBM} was proposed,  in which the time-frequency bins where  speech is active are marked  `1', and the other bins are set to  `0'. Here, the \ac{IBM} is estimated from the noisy input.  The noisy signal is then multiplied by the \ac{IBM} to reduce the noisy bins. This \emph{hard decision} approach is however not satisfactory for speech enhancement. This led to the development of an \ac{IRM}, which applied soft enhancement and had better results.
	
	These  \ac{DNN} approaches have several drawbacks. First, their fully-connected architecture has to deal with the massive variability of  speech in the input. Second, these approaches need to be trained on huge databases with varying noises in order to be able to minimize the unfamiliar noise \cite{wang2015large}. Finally, even if trained on a large database, there can   still be a mismatch between the test phase and the training phase, since  noise in real-life scenarios is always novel.
	
	To overcome these hurdles, a new phoneme-based architecture was introduced with an \ac{ASR} system \cite{wang2016phoneme}. In this architecture, a set of \ac{DNN}s were trained separately, one for each phoneme, with its own database to find the \ac{IRM}. Given a new noisy input, the \ac{ASR} system outputs the index of the phoneme associated with the current input, and that phoneme \ac{DNN} is activated to find the \ac{IRM}. This approach improved performance in terms of noise reduction and more accurate \ac{IRM} estimation. However,  when the \ac{ASR} system is incorrect, the wrong \ac{DNN} is activated. Additionally, the continuity of the speech is disrupted by mistakes in the \ac{ASR} system. Finally, the \ac{ASR} was not part of the training phase.
	
	Chazan et al.  \cite{chazan2016iwaenc} presented a similar architecture, but instead of the \ac{ASR} system, another \ac{DNN} was used as a phoneme-classifier. This approach produced better performance than the fully-connected approach and the phoneme-based architecture with \ac{ASR} without joint training. Yet, this algorithm still had major drawbacks. First, a phoneme-labeled database is essential, a requirement which is not always tractable. Second, the phoneme-classifier was not part of the joint training, which might prevent the training from producing even better results. Finally, the \ac{MSE} loss function,  used to train the network is not the natural choice for training binary classification tasks such as finding the \ac{SPP}. The approach presented in this study can be viewed as an extension of the method in Chazan et el.   \cite{chazan2016iwaenc},  and our goal is to present an enhancement approach that overcomes  the problems describe above. In the next section we described the previous work  \cite{chazan2016iwaenc}  in  more details.
	
      The \ac{MoE} approach, which was introduced more than twenty years ago \cite{mixture_of_experts,jordan1994hierarchical}, is based on the
principle of divide and conquer.
\ac{MoE} combines the decisions of several `experts', each of
which specializes in a different part of the input space.
The model  has three main components:
several experts that are either regression functions or
classifiers, a gate that makes soft partitions of the input space
and defines those regions where the individual expert opinions
are trustworthy, and  a weighted sum of experts,
where the weights are the input-dependent gates. The MoE model allows the individual experts to specialize on
smaller parts of a larger problem, and it uses soft partitions
of the data implemented by the gate. Previous work on \ac{MoE} has  focused on
different facets including using different types of expert models such as SVMs \cite{collobert2002parallel} and  Gaussian processes \cite{tresp2000mixtures}.
 A comprehensive  survey of  \ac{MoE} theory and applications can be found in \cite{yuksel2012twenty}.
 In spite of the huge success of deep learning there are not many studies that have explicitly utilized and analyzed \acp{MoE} as an  architecture component of a  neural network.   Eigen et al. \cite{Eigen} suggested the  extension of MoE to a deep model by stacking two layers of a mixture of experts (where each expert is a feed forward network) followed by a softmax layer.

In this work, we present  \ac{DMoE} modeling  for speech enhancement.
The noisy speech signal contains several different regimes which have different relationships between the input and the output based for instance, on phoneme identity or the coarser distinction between voiced and unvoiced phonemes. The proposed enhancement approach is based on a  \ac{DMoE} network where the   experts are DNNs, which are combined  by a  gating DNN. Each expert is responsible for enhancement in a single speech regime and the  gating network finds the suitable regime in each time frame.
  Each expert estimates an \ac{SPP} and  the  local \ac{SPP} decisions are averaged, based on the gating function, into a  final \ac{SPP} result. In our approach there is no need for  phoneme-labeled data, since the gating DNN splits  the input space into sub-areas in an unsupervised manner.

	The rest of the paper is organized as follows. In Sec~\ref{sec:problem} the problem formulation and previous work are presented. Sec.~\ref{sec:modealgorithm} introduces the new \ac{DMoE} model, and Sec. \ref{sec:dmoedforen} describes the application of \ac{DMoE} modeling to speech enhancement.
  The comprehensive experimental results using speech databases with various noise types are presented in Section~\ref{sec:experiments}. Sec~\ref{sec:disccusion} discuses the attributes of the  algorithm.  Finally, some conclusions are drawn and the paper is summarized in Section~\ref{sec:summery}.

	\section{Problem formulation and prior art}\label{sec:problem}
	
	\subsection{Problem formulation}
	Let $s(t)$ denote a sample of speech signal at time $t$. Let $x(t)=s(t)+n(t)$ denote the observed noisy signal where additive noise $n(t)$ was added to the clean speech.
	
	The \ac{STFT}  with a frame of length $L$ of $x(t)$ is denoted by $X_n(k)$, where $n$ is the frame index and $k=0,1,\hdots,L-1$ denotes the frequency band index. Similarly, $S_n(k)$ and $N_n(k)$ denote the \ac{STFT} of the speech and the noise only, respectively.
	
	Define the log-spectrum of the noisy signal at a single time frame by $\x$, such that the $k$-th component is  $x_{k}=\Log|X(k)|$ where $k=0,\ldots,L/2$. Note, that the other frequencies  can be obtained by the symmetry of the \ac{DFT}.  Similarly, $\s$ and $\n$ denote the log-spectrum of the corresponding speech and the noise only signals, respectively.
	
	N{\'a}das et al. \cite{6}  suggested  a maximization approximation modeling, in which  each time-frequency bin of the noisy signal log-spectrum, $x_k$, is dominated by the maximum between the log-spectrums of the speech, $s_k$ and the noise, $n_k$,
	\begin{equation}
	x_k=\max(s_k,n_k), \hspace{1cm} k=0,\ldots,L/2.
	\label{max}
	\end{equation}
	This approximation was utilized and yielded high performance in speech recognition \cite{6}, speech enhancement \cite{3, 23, chazan2016hybrid,chazan2016iwaenc} and speech separation \cite{26,29}.
		Following this approximation, a binary mask can be built for each $k$-th frequency,
	\begin{equation}
	b_{k}=\begin{cases}
	1 &s_{k}>n_{k}\\
	0 &s_{k}\leq n_{k}
	\end{cases}.
	\label{binary}
	\end{equation}
	We denote the binary-mask vector of all the frequencies at a given time frame as $\boldsymbol{b}=\left[ {b_0,\ldots,b_{L/2}}\right] $.

	In the enhancement task only the noisy signal $\x$ is observed, and the goal is to find an estimation $\hat{\s}$ of the clean speech $\s$. If the true binary mask $\boldsymbol{b}$ had been available, a soft attenuation algorithm could have been used,
	\begin{equation}
	\hat{\s}= \boldsymbol{b}  \cdot  \x+(1-\boldsymbol{b}) \cdot (\x-\beta),
	\label{soft_enh}
	\end{equation}
	where $\beta$ is the attenuation level.
	However, the binary-mask $\boldsymbol{b}$ is not available for the enhancement procedure. Instead, we need to compute the  \ac{SPP} vector
 $ \rrho=\left[ {\rho_0,\ldots,\rho_{L/2}}\right] $,
 from the noisy signal:
	\begin{equation}
	\rho_k=p(b_k=1|\x)=p(s_k>n_k|\x).
	\label{rho}
	\end{equation}
	The goal of this work is to build a DNN with a task-specific architecture  tailored to the problem of
 finding the SPP $\rrho$.
 Then together with (\ref{soft_enh}) the speech is enhanced:
	\begin{equation}
	\hat{\s}= \rrho  \cdot  \x+(1-\rrho) \cdot (\x-\beta).
	\label{final_soft_enh}
	\end{equation}
	
	Training is based on a labeled training dataset that consists of frames (in the log-spectrum) of noisy signals along with the associated true binary masks. 	In principle, we can train  a \ac{DNN} with a standard architecture  based on a  pipeline of fully-connected layers.  Several studies \cite{wang2015large,1} have shown that this approach is beneficial for  speech enhancement in terms of noise reduction.
	
	However, there are some drawbacks to using a fully-connected network architecture. The input to the \ac{DNN} is highly non-homogeneous. Speech is composed of different phonemes and  its structure varies over time. Thus, it is a difficult task to  train  a single \ac{DNN} to deal with this variability. It is also  a very difficult task for a single network to preserve the harmony structure  of the speech.
A single network also has  to address different noise types. Since there are many types of noise there can be  a mismatch between the train and test conditions when  an unfamiliar noise is introduced at test time. This leads to a decline in  performance. One possible strategy to overcome this problem is to use a massive database that consists of many hours of recorded data combined with many noise types \cite{wang2015large}. This approach, however, leads to a long training phase \cite{wang2015large}.  Therefore, a standard fully-connected  \ac{DNN}  might not be the best solution for the speech enhancement task.
	
	\subsection{Prior art based on phoneme-labeled data}
	
To overcome the problems described above,  the speech frames can be grouped such that the variability within each group is not high. For example, we can split the noisy training data according to the phoneme labels. Then, it might be beneficial to train an SPP estimation network separately on each homogeneous subset of the input.

	In a previous work  \cite{chazan2016iwaenc}, a phoneme-based architecture was suggested to find the \ac{SPP}. In this procedure, separate \ac{DNN} is allocated  for each phoneme to find the SPP given the phoneme class information. Additionally, a phoneme-classifier  produce the phonemes' posterior distribution. Finlay, the  \ac{SPP}  found by a weighted average of all the phoneme-based SPP estimations.
	The network is  trained \textit{jointly} to minimize the loss function which was chosen to be the \ac{MSE} between the final-\ac{SPP}, $\rrho$ and the binary-mask, $\boldsymbol{b}$.
 This training takes place in two steps. First, a pre-training stage done uses phoneme-labeled data. In this step each component of the architecture is separately trained to obtain a parameter initialization for the network.  In the second stage all phoneme-based sub-networks are  jointly trained.  This approach was shown to yield better results compared to the widely used fully connected architecture.
	Nevertheless, this approach still has some drawbacks. First, a phoneme-labeled database is essential for  phoneme-based training. This information, however, is not always available. Second, it is not obvious that dividing the speech signals according to the uttered phoneme  is the best strategy. Sometimes it is more worthwhile to  let the network automatically split the data in a way that   best suits enhancement.

		\section{A Deep Mixture of Experts}\label{sec:modealgorithm}
	In this section  we present the  Deep Mixture of Experts (DMoE) framework, and in the next section we apply it to  speech enhancement.
	
The Mixture-of-Experts (MoE) model introduced by Jacobs et al. \cite{mixture_of_experts,jordan1994hierarchical},  provides important paradigms for learning a classifier from data.
The objective of this framework is to describe the behavior of certain phenomena, under the assumption that there are separate processes involved in the generation of the data under analysis. The use of MoE makes it possible to combine many simple models to generate a more powerful one.
 The main idea is based on the  `divide-and-conquer' principle that is often used to attack a complex problem by dividing it into simpler problems whose solutions can be combined to yield a solution to the complex problem.

 The MoE model  is comprised of a set of classifiers that perform the role of experts, and a set of mixing weights determined by a gating function
 that selects the appropriate expert.   The experts are responsible for modeling the generation of outputs, given a certain condition on the input space, and are combined by a set of local mixing weights determined by the gating function, which depends on the input.

  We can view the MoE model as a two step process that produces a decision $y$ given an input feature set $x$. We first use the gating function to select an expert and then  apply the expert to determine the output label. The index of the selected expert can be viewed as an intermediate  hidden random variable denoted by $z$.
 Formally, the MoE conditional distribution  can be written as follows:
\begin{equation}
p(y|x;\theta) = \sum_{i=1}^m p(z=i|x;\theta_g) p(y|z=i,x;\theta_i)
\label{likmoe}
\end{equation}
such that $x$ is the feature vector, $y$ is the classification result, $z$ is a hidden random variable that selects the expert that is applied and $m$ is the number of experts.
The model parameter-set $\theta$ is composed of   the parameter-set of the gating function  $\theta_g$ and    parameter-sets $\theta_1,\dots,\theta_m$  for  the $m$ experts.

\begin{figure}
		\centering
		\includegraphics[trim=100 120 200 50 ,clip, scale=0.45]{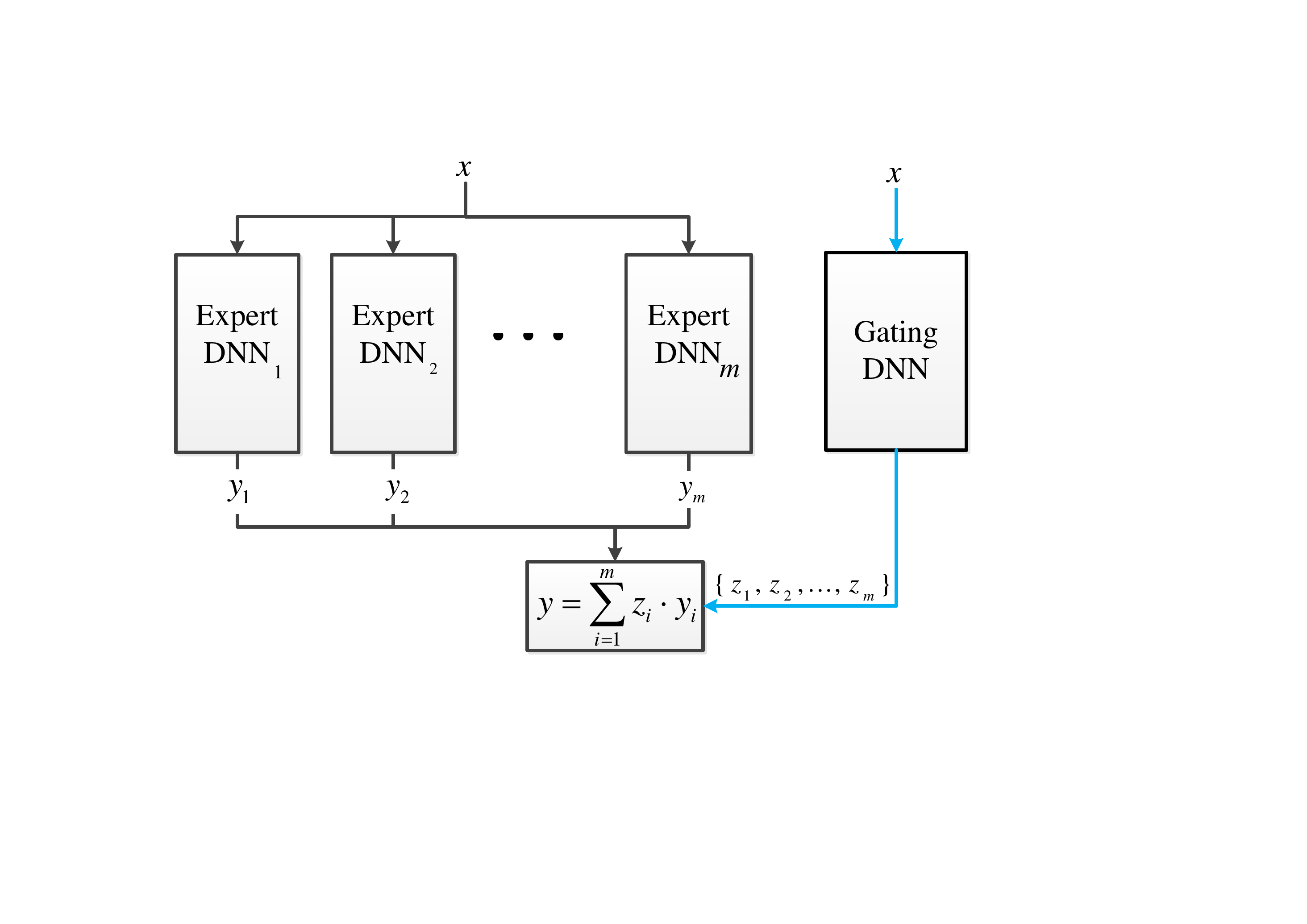}
		\caption{Deep Mixture of Experts (DMoE) architecture.}
		\label{fig:block_diagram}
	\end{figure}

 In the speech enhancement task $x$ is the log-spectrum vector of the noisy speech, $y$ is the  binary mask information and $z$ is a hidden speech state; e.g., the phoneme identity or a voiced/unvioiced indication. (Note that in the other sections of this paper we denote the binary mask information as $\boldsymbol{b}$.)
  In that case the gating function in the enhancement procedure estimates the phoneme given the noisy speech and each expert is associated with a phoneme and is responsible for enhancement of a noisy phoneme utterance.
The \ac{DMoE} is illustrated in Figure \ref{fig:block_diagram}.

We next address the problem of learning the MoE parameters (i.e. the parameters of the experts and the gating function) given a training dataset
  $(x_1,y_1),\dots,(x_n,y_n)$.
  The likelihood function of the MoE model parameters is:
 \begin{equation}
  L(\theta) = L(\theta_g, \theta_1,...,\theta_m) = \sum_t  \log p ( y_t|x_t; \theta).
  \label{likmoee}
  \end{equation}
  Since the selected expert used to produce $y_t$ from the feature set $x_t$ (i.e. the value of the r.v. $z_t$) is hidden, it is natural to apply the EM algorithm to find the maximum-likelihood parameters \cite{jordan1994hierarchical}.
    The EM auxiliary function is:
  \begin{equation}
  Q(\theta,\theta^0) = E_{p(z|x,y;\theta^0)}( \log p (y,z|x;\theta))
  \end{equation}
  such that $\theta^0$ is the current parameter estimation.
        In  the E-step we apply Bayes' rule to estimate the value of the selected expert based on the current parameter estimation:
\begin{equation}
w_{ti}= p(z_t=i|x_t,y_t;\theta^0) = \frac { p(y_t|x_t,z_t=i)p(z_t=i|x_t)}{p(y_t|x_t)}.
\label{estep}
\end{equation}
 The M-step decouples the parameter estimation of the different components of the MoE model. We can   optimize each of the experts and the gating function separately. The updated parameters of the gating function are obtained by  maximizing the weighted likelihood function:
 \begin{equation}
 L(\theta_g) = \sum_t \sum_i w_{ti} \log  p(z_t=i|x_t; \theta_g)
 \label{emm1}
\end{equation}
 and the updated parameters of the $i$-th expert are obtained by  maximizing the function:
  \begin{equation}
 L(\theta_i) = \sum_t w_{ti} \log  p(y_t|x_t, z_t=i; \theta_i)
 \label{emm2}
\end{equation}
The general EM  theory guarantees a monotone  convergence of the model-parameter estimations (to a local maximum).

In the case where both the experts and the gating functions are implemented by DNNs, we denote this modal Deep Mixture of Experts (DMoE).
The compound DNN model is expressed as follows:
\begin{equation}
p_{\mbox{\tiny NN}}(y|x;\theta) = \sum_i p_{\mbox{\tiny NN}}(z=i|x;\theta_g) p_{\mbox{\tiny NN}}(y|z=i,x;\theta_i).
\label{likdmoe}
\end{equation}
In this model $\theta_g$ is the parameter-set of the DNN that implements the gating function and  $\theta_i$ is the parameter-set of the
DNN that implements the $i$-th expert.  We can still apply the EM algorithm described above to train the a DMoE.
In this case in the M-step we need to train the experts and gating neural networks using the cross-entropy cost function defined by Eq. (\ref{emm1}) and (\ref{emm2}). We can thus iterate between EM steps and DNN training.

There are, however, several drawbacks to using the EM algorithm to train a DMoE. In  each M-step iteration we need to train a new DNN.
There is no closed-form solution for this non-concave  maximization task. The DNN learning is performed by a stochastic gradient ascent and there is no guarantee for monotone improvement of the likelihood score.  The EM algorithm is
a greedy optimization procedure that is notorious for getting stuck in local optima. In most EM applications there is a closed-form solution for the optimization performed at the M-step. Here, since we need to retrain the experts and gating  DNNs at each M-step  (\ref{emm1}) (\ref{emm2}),   even a monotone improvement of the likelihood  is not guaranteed.
The main problem of iterating between EM-steps and neural network training, however, is that it does not scale well.
 The framework
requires training a neural network in each iteration of the EM algorithm. For real-world, large-scale
networks, even a single training iteration is a non-trivial challenge.

 In this study we replace the EM algorithm, which  learns the expert and gating networks at each step separately, by a neural-network training procedure that  simultaneously  trains all the sub-networks by directly maximizing the likelihood function \begin{equation}
L(\theta)= \sum_{t=1}^n \log  (\sum_{i=1}^m p_\text{{\tiny NN}}(z_t=i|x_t;\theta_g) \cdot p_\text{{\tiny NN}}(y_t|z_t=i,x_t;\theta_i))
%L(\theta)= \sum_{t=1}^n \log  (\sum_{i=1}^m \underbrace{p(z_t=i|x_t;\theta_g)}_{p_i(t)} \cdot \underbrace{p(y_t|z_t=i,x_t;\theta_i))}_{p_{y,i}(t)}
\end{equation}
In this architecture, the experts  and the gating networks are  components of a single  network and are simultaneously trained  with the same objective function. In order to update the network parameters we apply a back-propagation algorithm. It can be easily verified that the back propagation equation for the parameter set  of the $i$-th expert is:
\begin{equation}
\frac{\partial L}{\partial \boldsymbol{\ttheta}_{i}} = \sum_t w_{ti} \cdot \frac{\partial} {\partial \ttheta_{i}}
 \log   p_\text{{\tiny NN}}(y_t|z_t=i,x_t;\theta_i)
\label{dmoefder1}
\end{equation}
such that $w_{ti}$ is the posterior distribution of the gating random variable:
\begin{equation}
\begin{aligned}
w_{ti}= &p_\text{{\tiny NN}}(z_t=i|x_t,y_t;\theta) =\\ &\frac { p_\text{{\tiny NN}}(y_t|x_t,z_t=i;\theta_i)p_\text{{\tiny NN}}(z_t=i|x_t;\theta_g)}{p_\text{{\tiny NN}}(y_t|x_t;\theta)}.
\end{aligned}
\label{sestep}
\end{equation}
Note that this definition coincides with the E-step of the EM algorithm defined in Eq. (\ref{estep}).
In a similar way,  the  back-propagation equation for the parameter set of the gating DNN is:
\begin{equation}
\frac{\partial L}{\partial \boldsymbol{\ttheta}_{g}} = \sum_t  \sum_i w_{ti} \cdot \frac{\partial }{\partial \ttheta_g} \log  p_\text{{\tiny NN}}(z_t=i|x_t;\theta_g).
\label{dmoefder2}
\end{equation}
Note, that the back-propagation partial derivatives  (\ref{dmoefder1}) and (\ref{dmoefder2}) are exactly the  derivatives of the functions (\ref{emm1})  and  (\ref{emm2})  that are optimized by the M-step of the EM  algorithm. By training all the components of the  \ac{DMoE}  simultaneously we thus replace the two steps of the EM iterations by the single step of a gradient ascent optimization.

%
%Suppose we want to train a DNN soft classifier $p(y=i|x)$ a given training data $(x_1,y_1),\dots,(x_n,y_n)$.
%    	A fully connected architecture is very popular approach for DNN classifiers, due to its simplicity and high performance. Yet, in some cases utilizing this approach to difficult tasks is not the best strategy. A single network has to deal with a lot of variability in the input data.
%    The network architecture and training procedure we described above as an alternative to the EM algorithm,  also provides an alternative to
%    the standard network architecture based on fully-connected layers.

%\end{document}

	\section{Deep Mixture Experts for Speech Enhancement}\label{sec:dmoedforen}
	In this section we apply the \ac{DMoE} principle to a speech enhancement task and describe the network specifics and training procedure.
	
	\subsection{Network description}
The goal in a speech enhancement task is to find an accurate SPP, $\rrho$ from a given noisy signal $\x$ using the \ac{DMoE} model.
	All the $m$ experts in the proposed algorithm are implemented by  \acp{DNN} with the same structure. The input to each \ac{DNN} is the noisy log-spectrum frame together with context frames. The network consists of 3 fully connected hidden layers with  500 \ac{ReLU} neurons each.

	The output layer that provides the \ac{SPP} binary decisions is composed of  $L/2$ sigmoid neurons, one for each frequency band. The SPP decision of the $i$-th expert on the $k$-th frequency bin is:
	\begin{equation}
		\rho_{i,k}=p(b_k=1|\x;\theta_i).
		\label{rho_ki}
	\end{equation}
Let $h_i(\x)$ be the value of the final hidden layer of the $i$-th expert. The SPP prediction can be written as
  $\rho_{i,k} = \sigma ( w_{ik}h_i(\x)+a_{ik})$
  such that $\sigma$ is the sigmoid transfer function and  $w_{ik},a_{ik}$ are the parameters of the affine input function to the sigmoid neuron. The fact that all the frequency bands are simultaneously  estimated from $h_i(\x)$ enables the network to reconstruct the harmonic structure.

  Although the standard  \ac{MoE} approach uses the same input features for both the experts and the gating networks, here the log-spectrum of the noisy signal, $\x$, is utilized as the input for the experts alone, and the gating \ac{DNN} is fed with the corresponding \ac{MFCC} features denoted by $\vv$.
   MFCC, which  is based on frequency bands,  is a more compact representation than a linearly spaced log-spectrum  and  this frequency warping is known  for its better representation of sound classes \cite{hermansky2013ASRproperties}. We  found that using the MFCC representation for the gating DNN both slightly improves performance and significantly reduces the input size.   
   
   The architecture of the gating DNN is also composed of 3 fully connected hidden layers with 500 \ac{ReLU} neurons each.  The output layer here is a  softmax function that produces the gating distribution on the $m$ experts.  The gating procedure therefore is:
	\begin{equation}
		p_i=p(z=i|\vv;\theta_{g}).
		\label{p_i}
	\end{equation}
The final \ac{SPP} is obtained by a weighted average  of the deep  experts' decisions:
	\begin{equation}
		\rrho=\sum_{i=1}^{m}p_i\cdot \rrho_i.
		\label{final_spp}
	\end{equation}

Given the SPP vector $\rrho$, the enhanced signal is finally obtained using Eq. (\ref{final_soft_enh}).
The suggested \ac{DMoE} algorithm for speech enhancement is presented in Algorithm \ref{algo:test}.

 The network was implemented in Keras \cite{chollet2015keras} on top of Theano backend \cite{theano} with ADAM optimizer \cite{adam}. To overcome the mismatch between the training and the test conditions, each utterance was normalized prior to the training of the network, such that the sample-mean and sample-variance of the utterance were  zero and one, respectively~\cite{12}. 
 In order to circumvent over-fitting of the DNNs to the training database, we first applied the \ac{CMVN} procedure to the input, prior to the training and test phases \cite{12}. Additionally, the dropout method~\cite{dropout} was  utilized on each layer. Finally, the batch-normalization method was applied to train acceleration on each layer \cite{batchnorm}.

\subsection{Training the \ac{DMoE} for speech enhancement}
To train the network we need to collect a  dataset of noisy speech and the corresponding binary  vectors.
	 Unlike \cite{chazan2016iwaenc}, here  neither a phoneme-labeled database nor pre-training are needed. Clean speech signals were contaminated with a \emph{single} noise type in a pre-defined \ac{SNR}. The log-spectrum of the speech and the noise are known and a binary mask based on the maximization approximation (\ref{binary}) was then computed.  Additionally, the corresponding \ac{MFCC} features were calculated. Finally, in order to  enhance the current frame of the noisy signal, context frame information   is known to provide better performance; therefore, each input contained four context frames from the past and four from the future.

Assume the training set is  $(\x_1,\xbb_1),...(\x_N,\xbb_N)$  such that $\x_t$ is a log-spectrum of  noisy speech and $\xbb_t$ is the corresponding binary
mask vector, and $N$ is the length of the database.	The training procedure aims to optimize the following log-likelihood function:.
	\begin{equation}
	L(\ttheta) = \sum_{t=1}^N \log p ( \xbb_t |\x_t ; \theta)
	\label{loss}
	\end{equation}
such that
	\begin{equation}
p ( \xbb_t |\x_t ; \theta) =
\sum_{i=1}^m  p(z_t\!=\!i|\vv_t;\theta_g) \prod_k  p( b_t(k) |\x_t, z_t\!=\!i ; \theta_i)
	\label{losst}
	\end{equation}
with $\vv_t$,  the MFCC coefficients at time $t$.

	\begin{algorithm}[]
	\SetAlgoLined
	\SetKwInOut{Input}{Input}\SetKwInOut{Output}{Output}
	
	\Input{\begin{itemize}
			\item Noisy speech log-spectral vector $\x$.
			\item Corresponding MFCC feature vector  $\vv$.
			\item \ac{DMoE} model parameters $\ttheta=\{ \ttheta_1,\ldots,\ttheta_m,\ttheta_g\} $.
		\end{itemize}}
		\Output{Estimated  log-spectral vector of the clean speech $\hat{\s}$. }  \vspace{0.1cm}
		{\textbf{Experts' DNNs}:\\
		%	Set the phoneme-based {MoG} parameters using $(\x_1,i_1), \ldots, (\x_N,i_N)$. \\
			\begin{itemize}
%				\item Build a \ac{DNN} with sigmoid output.
				\item Compute \ac{SPP} decision for each expert $i\in\{1,\dots,m\} $ and for each frequency band $k$ (\ref{rho_ki}):
					$$\rho_{i,k}=p(b_k=1 |\x;\theta_i)$$
	%			\item Concatenate all the frequencies together and get $\rrho_i$
			\end{itemize}
		}
	
	{\textbf{Gating \ac{DNN}:}}
		\begin{itemize}
%			\item build a \ac{DNN} with softmax output
			\item Compute the gating distribution (\ref{p_i}):
			$$p_i=p(z=i|\vv;\theta_g)$$
		\end{itemize}
		
		\textbf{Avergung Experts' decsions:}
		\begin{itemize}
			\item Compute final SPP (\ref{final_spp})
			$$\rrho=\sum_{i=1}^{m}p_i\cdot \rrho_i$$
		\end{itemize}

		{\textbf{Enhancement}}: \\
		\begin{itemize}
			\item 	Estimate the clean speech (\ref{final_soft_enh}):
			$$\hat{\s}= \rrho  \cdot  \x+(1-\rrho) \cdot (\x-\beta).$$
		\end{itemize}

		\caption{\ac{DMoE} speech enhancement algorithm.} \label{algo:test}

	\end{algorithm}
		\begin{algorithm}[]
	\SetAlgoLined
	\SetKwInOut{Input}{Input}\SetKwInOut{Output}{Output}
	
	\Input{\begin{itemize}
			\item Noisy speech log-spectral vectors $\x_1,...,\x_N$
			\item Corresponding MFCC feature vectors  $\vv_1,...,\vv_N$
            \item Binary mask vectors  $\xbb_1,...,\xbb_N$
  		\end{itemize}}
		\Output{\ac{DMoE} model parameters $\ttheta=\{ \ttheta_1,\ldots,\ttheta_m,\ttheta_g\} $.}  \vspace{0.1cm}
	
			\textbf{Neural network training}

			 Train a DNN in order to maximize the objective function:
$$	L(\ttheta) = \sum_{t=1}^n \log (\sum_{i=1}^m  p(z_t\!=\!i|\vv_t;\theta_g)
 \prod_k  p( b_t(k) |\x_t, z_t\!=\!i ; \theta_i))
$$

		\caption{\ac{DMoE} training procedure.} \label{algo:train}
	\end{algorithm}

	The back-propagation partial derivative of the parameters of the $i$-th expert is:
	\begin{equation}
\frac{\partial L}{\partial \boldsymbol{\ttheta}_{i}} = \sum_t w_{ti} \cdot \frac{\partial} {\partial \ttheta_{i}}
\sum_k \log   p(b_t(k)|z_t=i,x_t;\theta_i)
\end{equation}
where $w_{ti}$ (see Eq. (\ref{estep})) is the posterior probability that the $i$-th  expert was used to produce the binary mask $\xbb_t$.
The partial derivative of the gating parameter appears in (\ref{dmoefder2}).

Another decision we need to make is choosing the number of experts. We found that moving from a single expert to two experts yielded a significant improvement, but adding more experts had little effect. Hence, utilizing  Occam's razor principle, we chose the simpler model and set $m=2$.
In Sec~\ref{subsec:num_of_experts} we  show  performance empirically as a function of the number of experts. We also show that when we train \ac{DMoE} with two experts,  the gating network tends to direct voiced frames to one expert  and unvoiced frames to the second expert.
The training procedure is  summarized in Algorithm \ref{algo:train}.

	\begin{table*}[thbp]
		\begin{center}
			\begin{tabular}{@{}lcc@{}}
				\toprule
				\multicolumn{3}{c}{\textbf{Train phase}} \\
				\midrule
				& \textbf{Database} &  \textbf{Details} \\
				\midrule
				\textbf{Supervised \ac{DMoE} model} & TIMIT (train set)	& speech-like noise, SNR=10 dB, phoneme labels    \\
				\textbf{DMoE} & TIMIT (train set) & speech-like noise, SNR=10 dB\\
				\toprule
				\multicolumn{3}{c}{\textbf{Test phase}} \\
				\midrule
				& \textbf{Database} &  \textbf{Details} \\
				\midrule
				\textbf{Speech} & TIMIT (test set), {WSJ}&\\
				\textbf{Noise} &	NOISEX-92  & White, Speech-like, Room, Car, Babble, Factory   \\
				\textbf{{SNR}} &	- & -5, 0, 5, 10, 15~dB  \\
				\textbf{Objective measurements}	& - &PESQ, Composite measure  \\
				\bottomrule
			\end{tabular}
		\end{center}
		\caption{Experimental setup.}
		\label{tabel:experiment_setup}
	\end{table*}

	\section{Experimental study}\label{sec:experiments}
	In this section we present a comparative experimental study. We first describe the experiment setup in Sec.~\ref{subsec:expsetup}. Objective quality measure results are then presented in Sec.~\ref{subsec:objectmeasurs}. Finally, the algorithm is tested on an untrained database in Sec.~\ref{subsec:notimit}.

	\subsection{Experiment setup}\label{subsec:expsetup}
	To test the proposed \ac{DMoE} algorithm we  contaminated the speech signals with several types of noise from the NOISEX-92 database~\cite{43}, namely \emph{Speech-like}, \emph{Babble}, \emph{Car}, \emph{Room}, \emph{AWGN} and \emph{Factory}.
	The noise was added to the clean signal drawn from the test set of the TIMIT database (24-speaker core test set), with  5 levels of \ac{SNR} at $-5$~dB, $0$~dB, $5$~dB, $10$~dB and $15$~dB chosen  to represent various real-life scenarios.  The algorithm was also tested on the  \ac{WSJ} database \cite{22} which was collected with a different recording setup.
	We  compared the proposed algorithm to the  \ac{OMLSA} algorithm~\cite{17} with the \ac{IMCRA} noise estimator~\cite{32} which is a state-of-the-art algorithm for single microphone speech enhancement. The default parameters of the \ac{OMLSA} were set according to~\cite{40}. Additionally, we compared the proposed \ac{DMoE} algorithm to two other DNN-based algorithms. The first DNN  has a fully-connected architecture and  can be viewed as a single-expert network. We denote this network the \ac{DSE}. The second DNN is a supervised phoneme-based \ac{DMoE} architecture  \cite{chazan2016iwaenc}.   The network has 39  experts where each expert is explicitly  associated with a specific phoneme and  training uses  the phoneme labeling available in the TIMIT dataset. We denoted this phoneme-based supervised network  by S-\ac{DMoE}. Each
expert component in the \ac{DSE} and S-\ac{DMoE} networks has the same network architecture as the expert components of the proposed  \ac{DMoE} model.

	\subsubsection*{Training Procedure}
		In order to carry out a fair comparison, all the DNN-based algorithms were trained with the same database. We used the TIMIT database \cite{TIMIT} train set for the training phase and the test set for the testing. Note, that the train and test sets of TIMIT do not overlap.
		Clean utterances were contaminated with \emph{Speech-like} noise with an \ac{SNR} $=10$~dB. Note, that unlike most DNN-based algorithms \cite{wang2015large}, we trained the \ac{DMoE} network   only on a single noise type with a single pre-defined \ac{SNR} value.
The speech diversity modeling provided by the expert-set was found to be rich enough to handle noise types that were not presented in the training phase.

	In order to evaluate the performance of the proposed speech enhancement algorithm, several objective and subjective measures were used. The standard \ac{PESQ} measure, which is known to have a high correlation with subjective score~\cite{13}, was used. Additionally,  the composite measure suggested by Hu and Loizou~\cite{7}, was implemented. The composite measure  weights  the \ac{LLR}, the  \ac{PESQ}  and the \ac{WSS}~\cite{48} to predict the \emph{rating} of the \ac{Cbak}, the \ac{Csig} and the \ac{Covl} performance. The rating was based on the 1-5 \ac{MOS} scale, and the clean speech signal had a  \ac{MOS} value of 4.5.
	
	Finally, we  also carried out informal listening tests {with approximately thirty listeners.}\footnote{Audio samples comparing the proposed \ac{DMoE} algorithm with the \ac{OMLSA}, the DSE  and the S-\ac{DMoE} can be found in \url{www.eng.biu.ac.il/gannot/speech-enhancement/mixture-of-deep-experts-speech-enhancement/}.}
	Table~\ref{tabel:experiment_setup} summarizes the experimental setup.

	\subsection{Objective quality measure results}\label{subsec:objectmeasurs}
	We first evaluated the objective results of the proposed \ac{DMoE} algorithm and compared it with the results obtained by the  state-of-the-art \ac{OMLSA} algorithm, the DSE and the S-\ac{DMoE} phoneme-based algorithm \cite{chazan2016iwaenc}. The test set was the  core test-set of the TIMIT database.
	
\begin{figure*}[tbhp]
			\centering
			\begin{subfigure}[b]{0.5\textwidth}
				%				\centering
				\includegraphics[width=\textwidth]{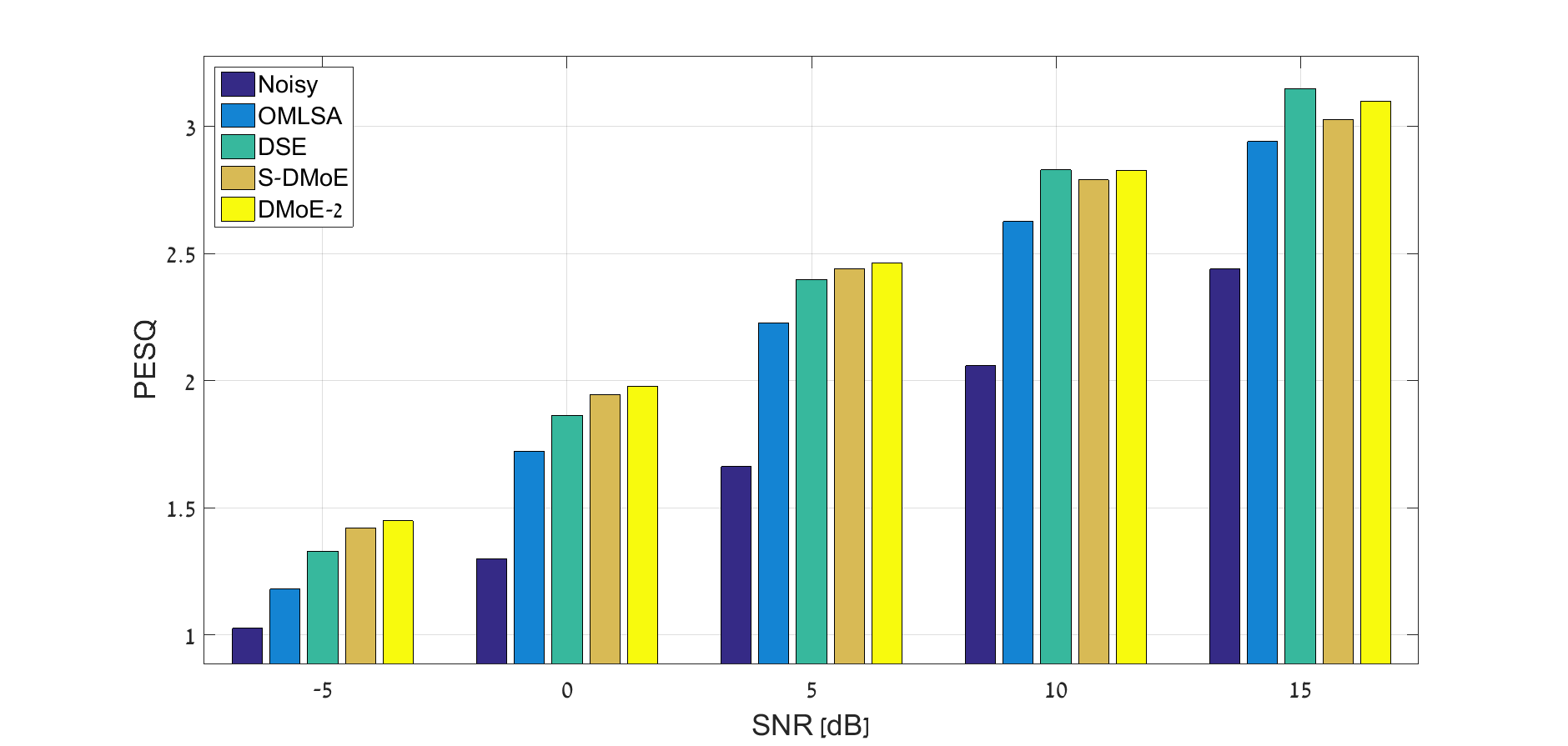}
				\caption{ {Speech} noise.}
				\label{fig:pesq_Speech}
			\end{subfigure}%
			\begin{subfigure}[b]{0.5\textwidth}
				%				\centering
				\includegraphics[width=\textwidth]{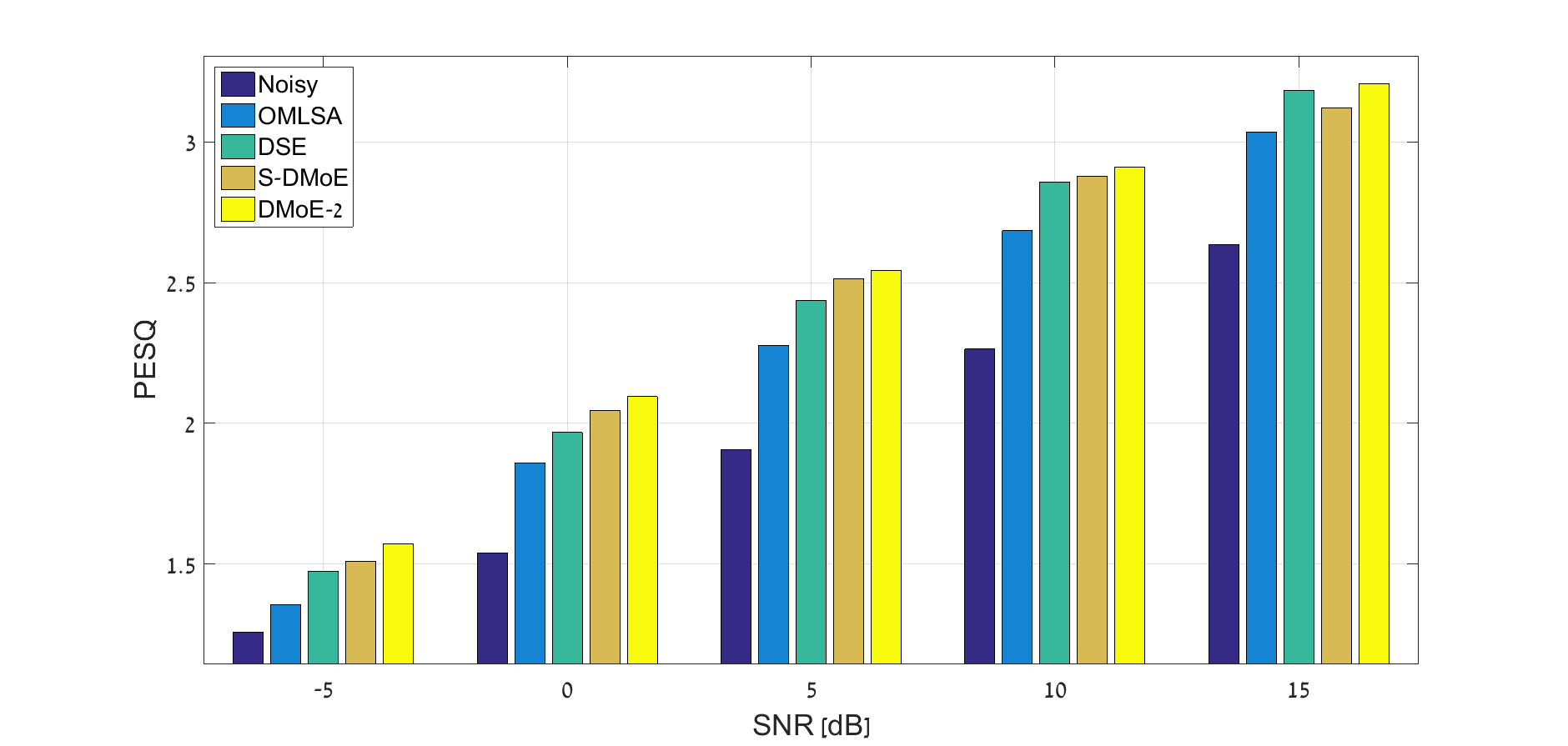}
				\caption{ {Room} noise.}
				\label{fig:pesq_Room}
			\end{subfigure}\\		
			\begin{subfigure}[b]{0.5\textwidth}
				\includegraphics[width=\textwidth]{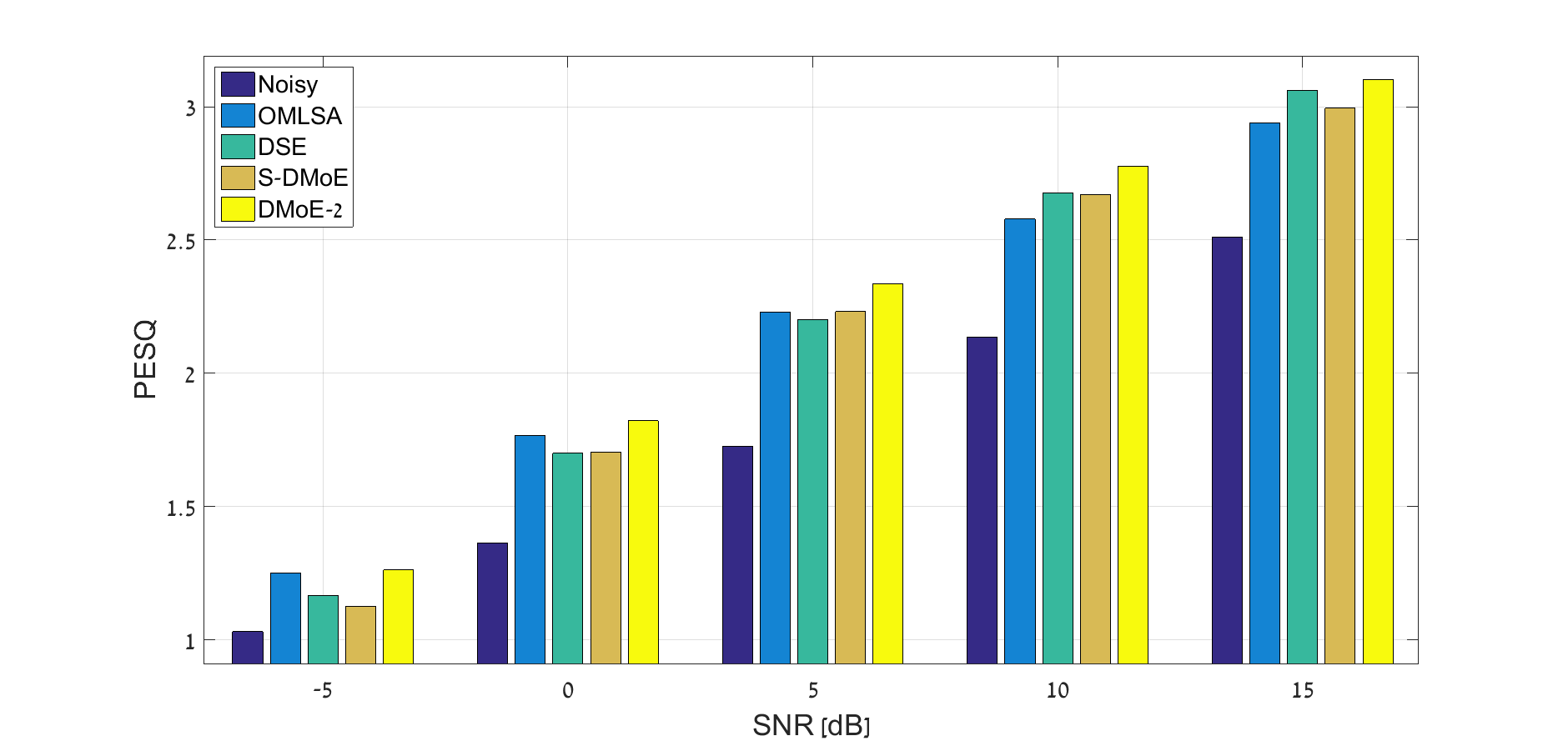}
				\caption{{Factory} noise.}
				\label{fig:pesq_factory}
			\end{subfigure}%
			\begin{subfigure}[b]{0.5\textwidth}
				%				\centering
				\includegraphics[width=\textwidth]{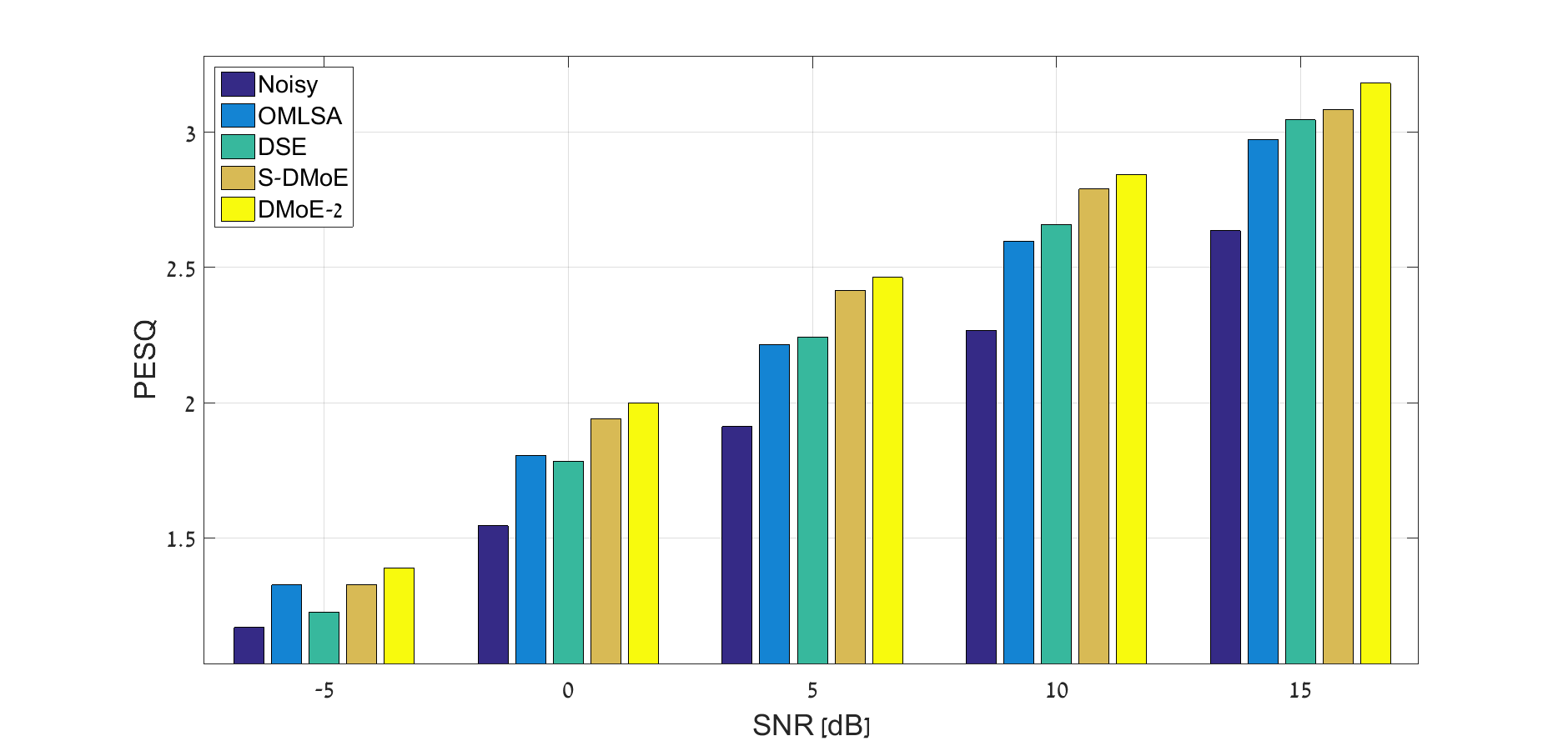}
				\caption{ {Babble} noise.}
				\label{fig:pesq_Babble}
			\end{subfigure}%
			\caption{Speech quality results (PESQ) on the TIMIT test-set for several noise types.}
			\label{fig:PESQ}
		\end{figure*}

\begin{figure*}[tbhp]
			\centering
			\begin{subfigure}[b]{0.5\textwidth}
				%				\centering
				\includegraphics[width=\textwidth]{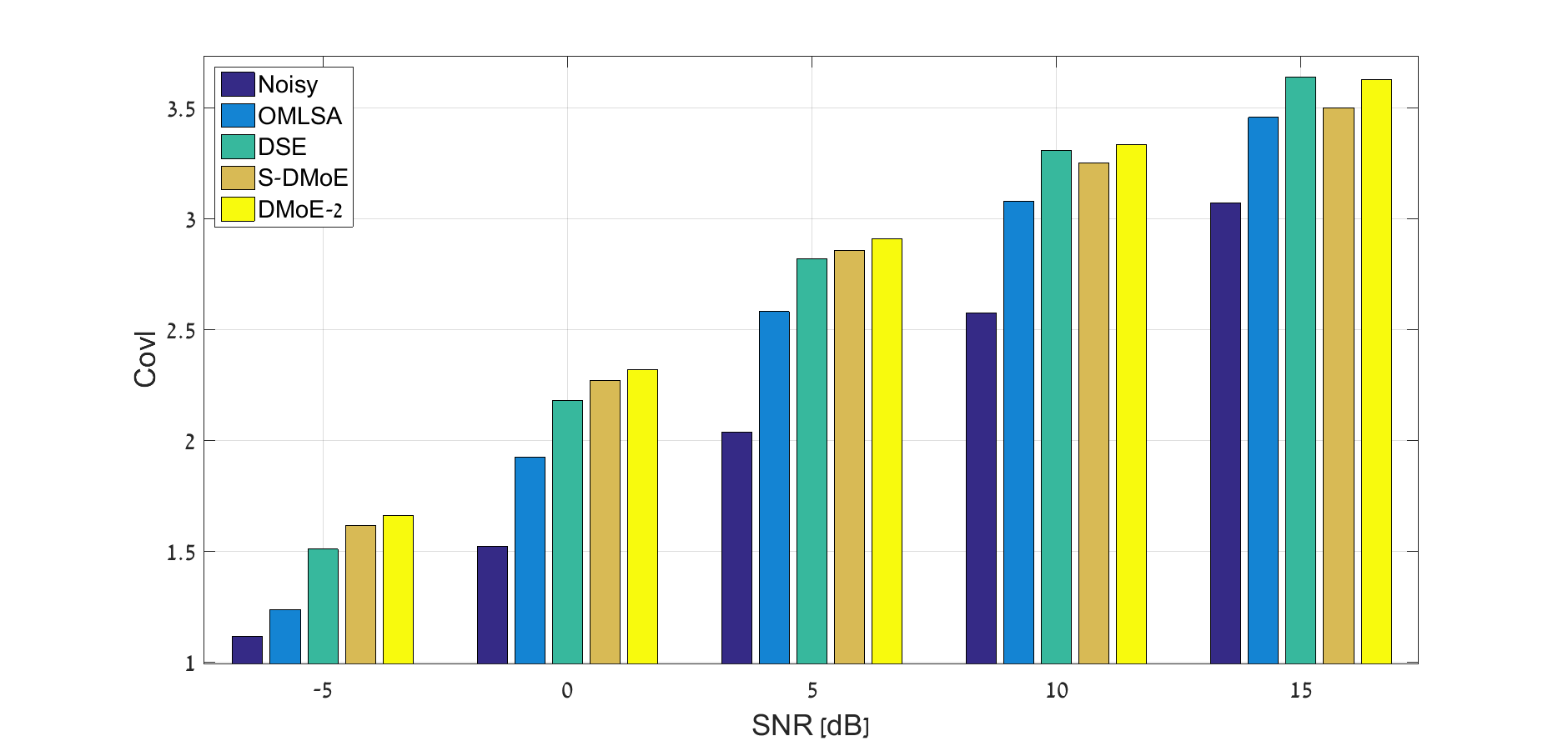}
				\caption{ {Speech} noise.}
				\label{fig:Speech}
			\end{subfigure}%
			\begin{subfigure}[b]{0.5\textwidth}
				%				\centering
				\includegraphics[width=\textwidth]{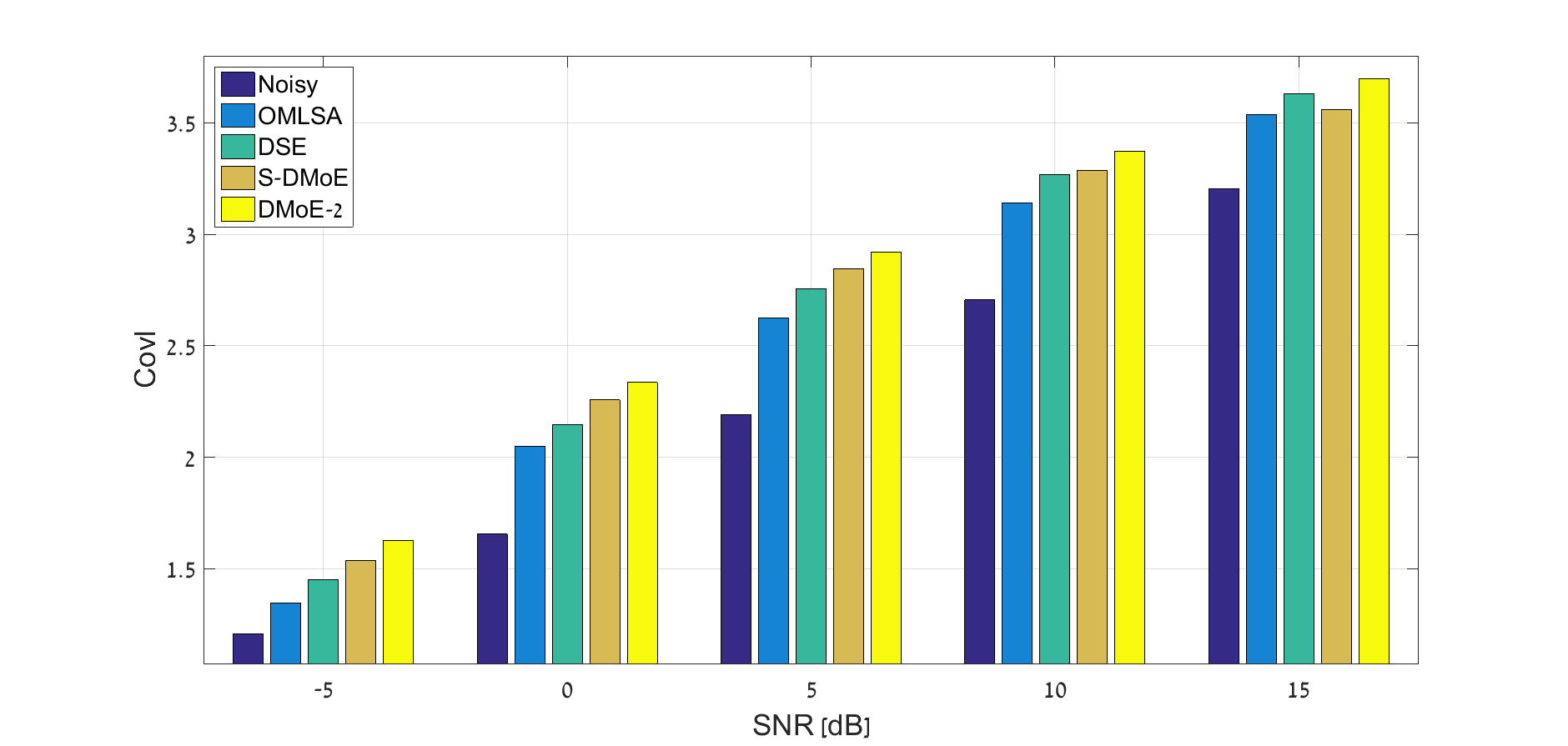}
				\caption{ {Room} noise.}
				\label{fig:Room}
			\end{subfigure}\\		
			\begin{subfigure}[b]{0.5\textwidth}
				\includegraphics[width=\textwidth]{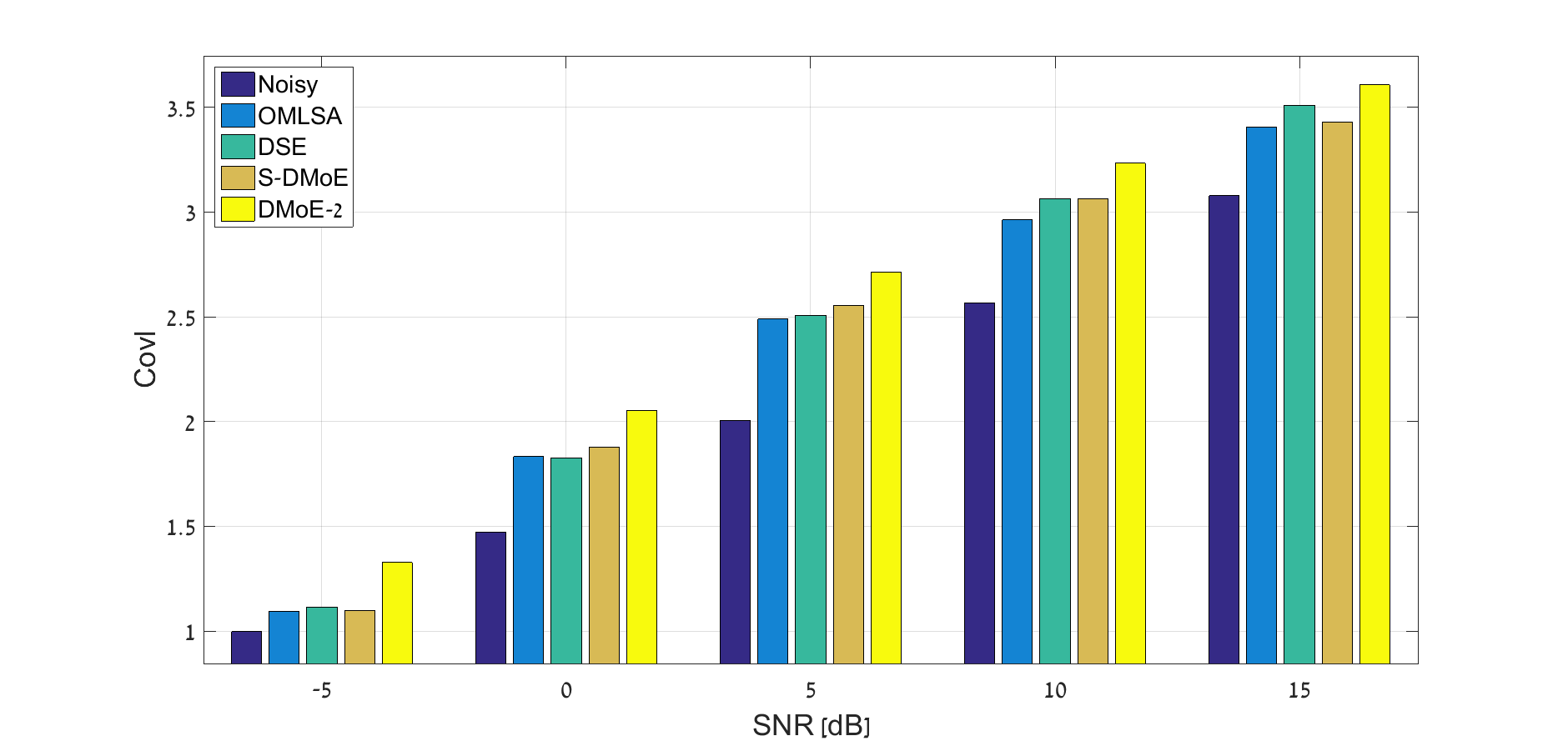}
				\caption{{Factory} noise.}
				\label{fig:factory}
			\end{subfigure}%
			\begin{subfigure}[b]{0.5\textwidth}
				%				\centering
				\includegraphics[width=\textwidth]{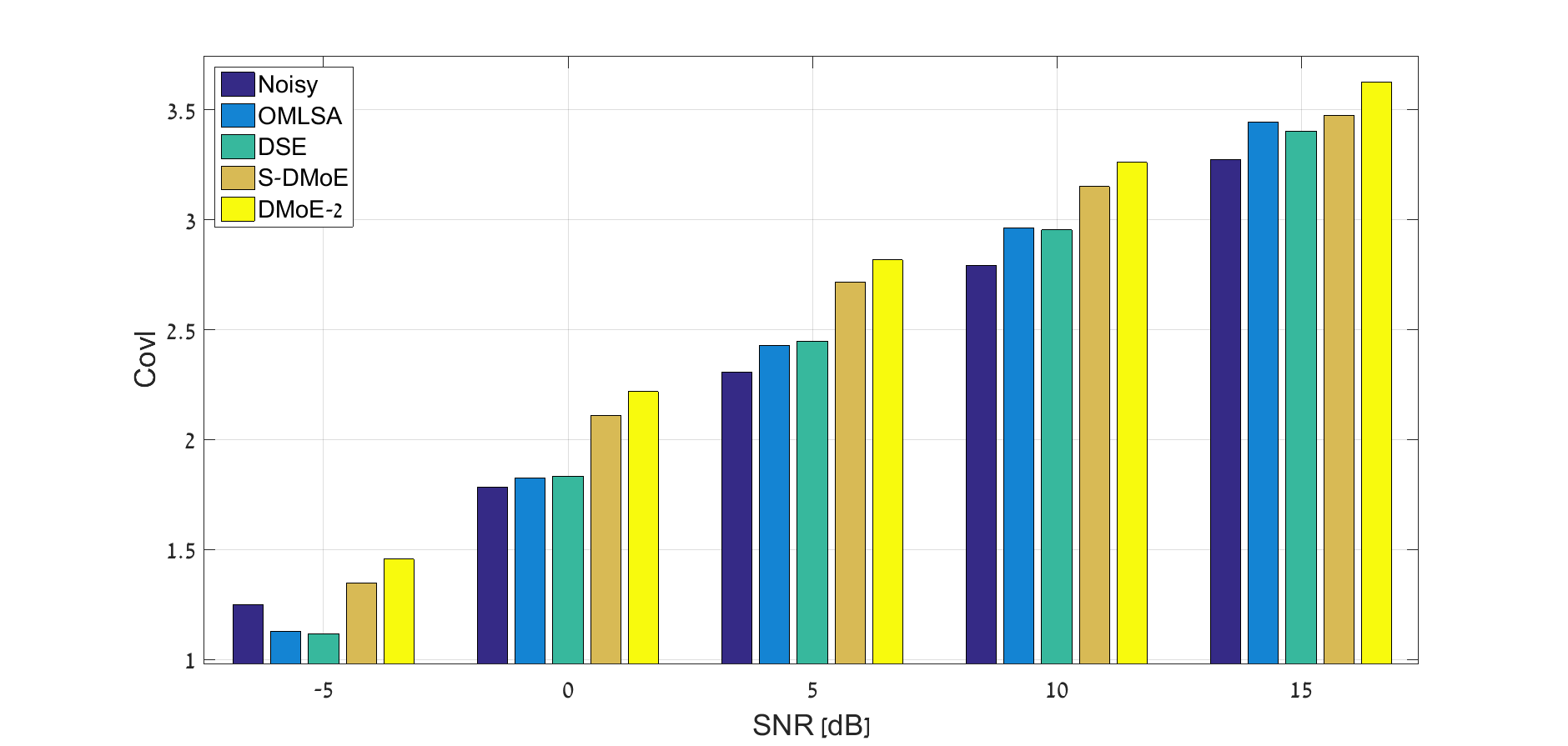}
				\caption{ {Babble} noise.}
				\label{fig:Babble}
			\end{subfigure}%
			\caption{Speech quality results (Covl) on the TIMIT test-set for several noise types.}
			\label{fig:Covl}
		\end{figure*}

	Fig.~\ref{fig:PESQ} depicts the \ac{PESQ} results for all algorithms for the {Speech-like}, {Room}, {Factory} and {Babble} noise types as a function of the input \ac{SNR}. In Fig.~\ref{fig:Covl} we show the Covl results for the same noises. %The results behave in a similar way,  with other noise types.
	
	It is evident that the proposed \ac{DMoE} algorithm outperformed the competing algorithms on the two objective measures.
	
\begin{figure*}[tbhp]
		\centering
		\begin{subfigure}[b]{0.5\textwidth}
			%				\centering
			\includegraphics[trim=0 170 0 200 , clip, width=\textwidth]{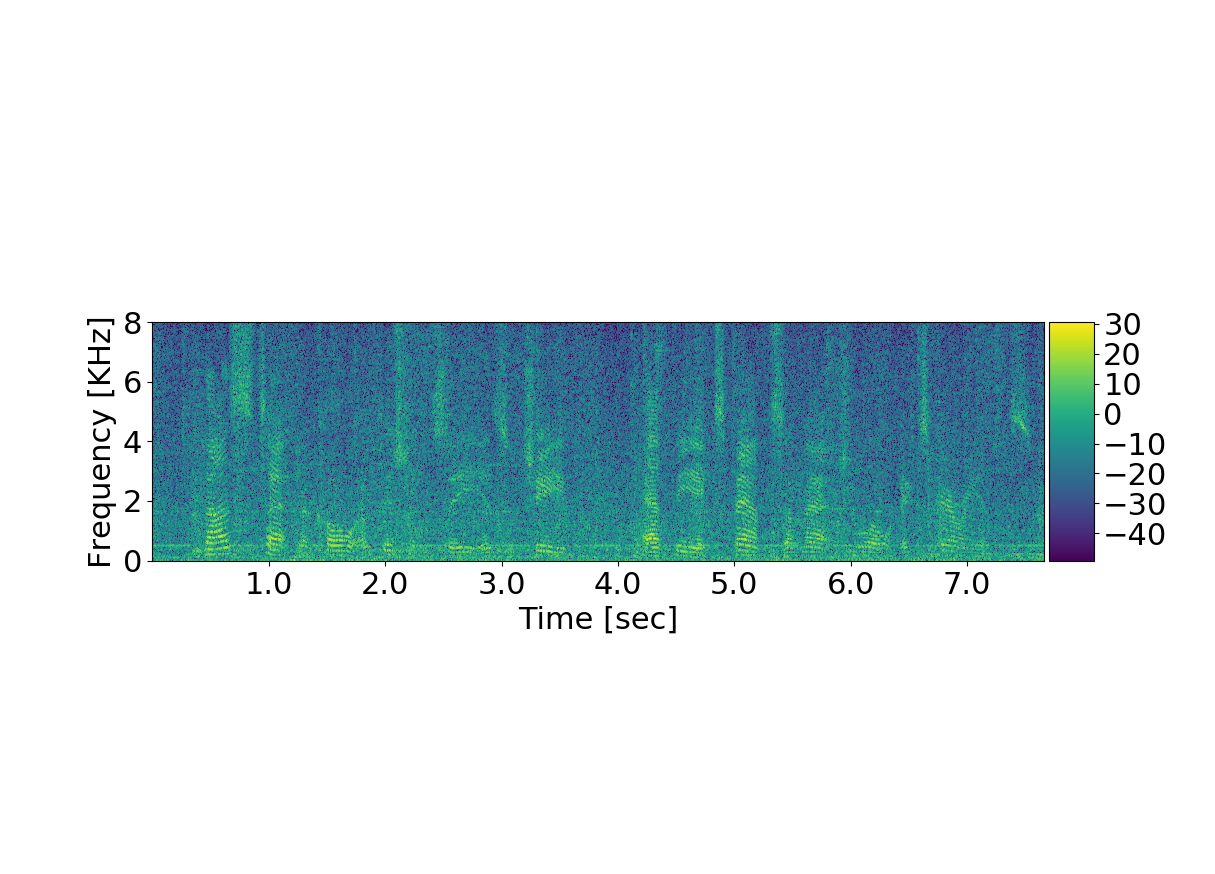}
			\caption{  Noisy.}
			\label{fig:spp_noisy}
		\end{subfigure}%
		\begin{subfigure}[b]{0.5\textwidth}
			%				\centering
			\includegraphics[trim=0 170 0 200 , clip , width=\textwidth]{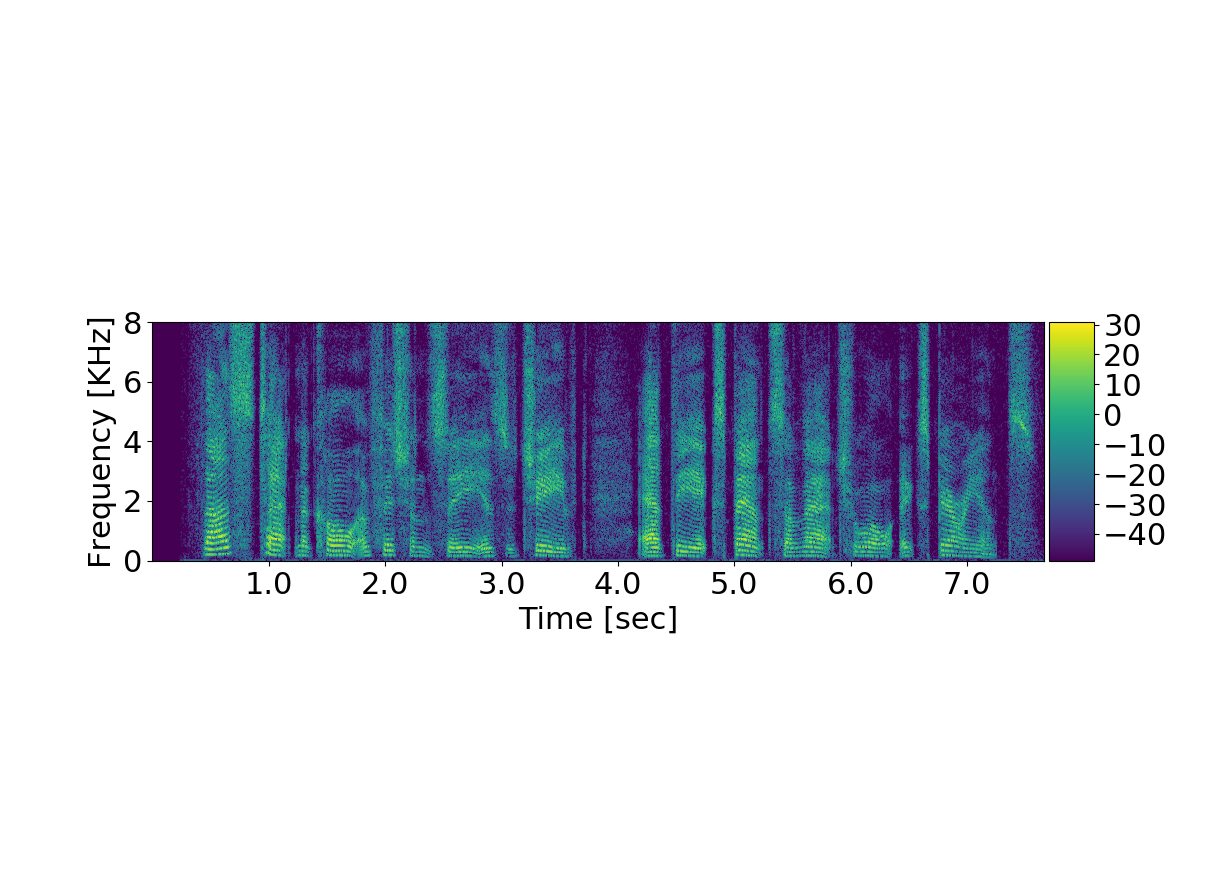}
			\caption{ Clean.}
			\label{fig:spp_clean}
		\end{subfigure}\\		
		\begin{subfigure}[b]{0.5\textwidth}
			\includegraphics[trim=0 170 0 200 , clip , width=\textwidth]{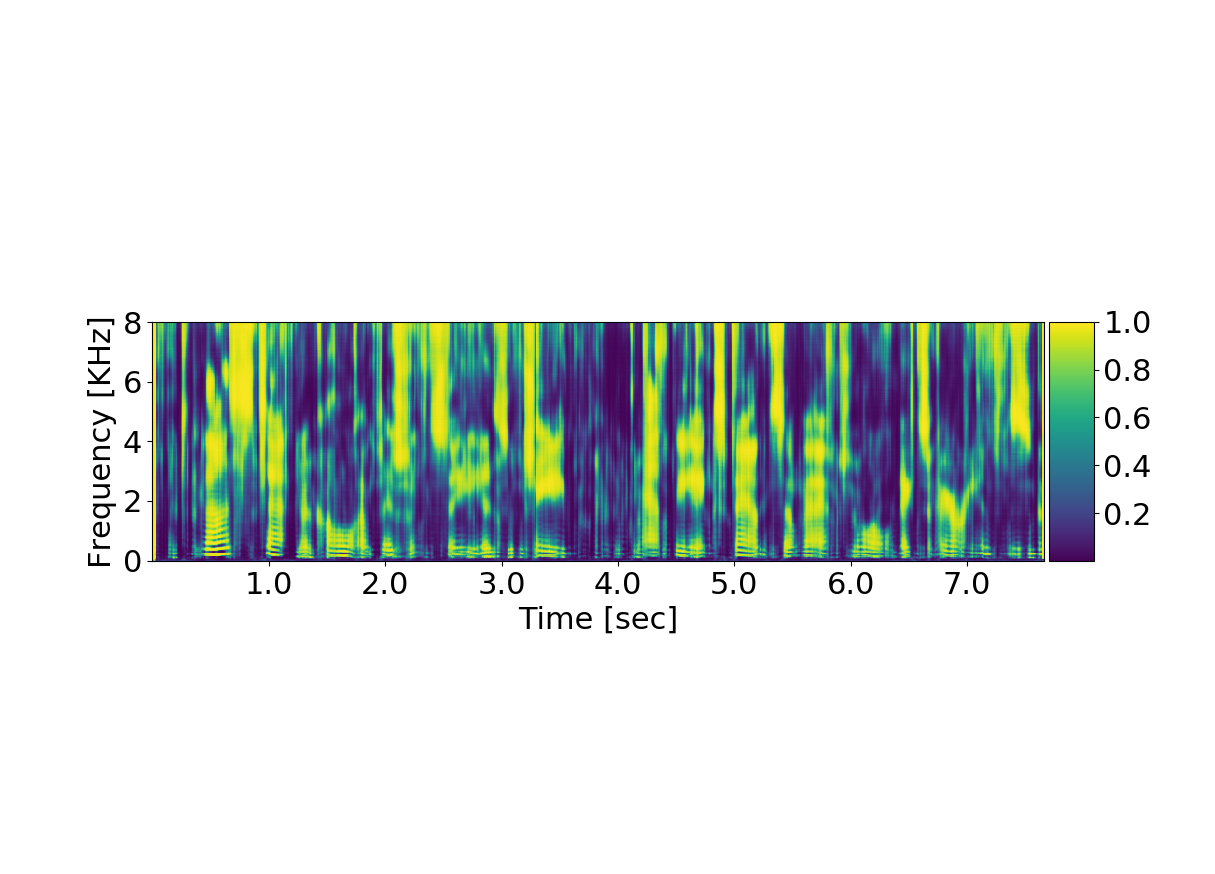}
			\caption{DSE SPP.}
			\label{fig:spp_sde}
		\end{subfigure}%
		\begin{subfigure}[b]{0.5\textwidth}
			%				\centering
			\includegraphics[trim=0 170 0 200 , clip , width=\textwidth]{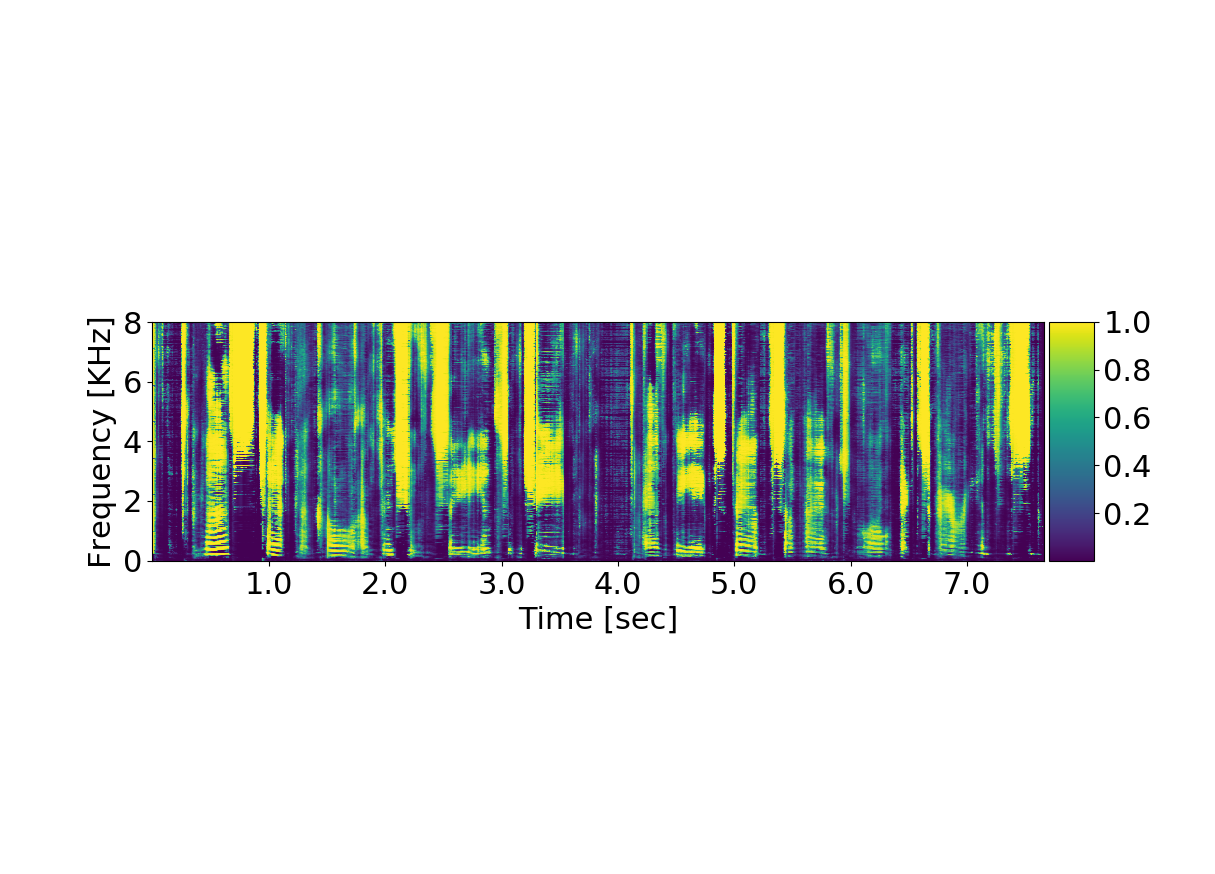}
			\caption{ S-DMoE SPP.}
			\label{fig:spp_s_dmoe}
		\end{subfigure}\\
		\begin{subfigure}[b]{0.5\textwidth}
			%				\centering
			\includegraphics[trim=0 170 0 200 , clip , width=\textwidth]{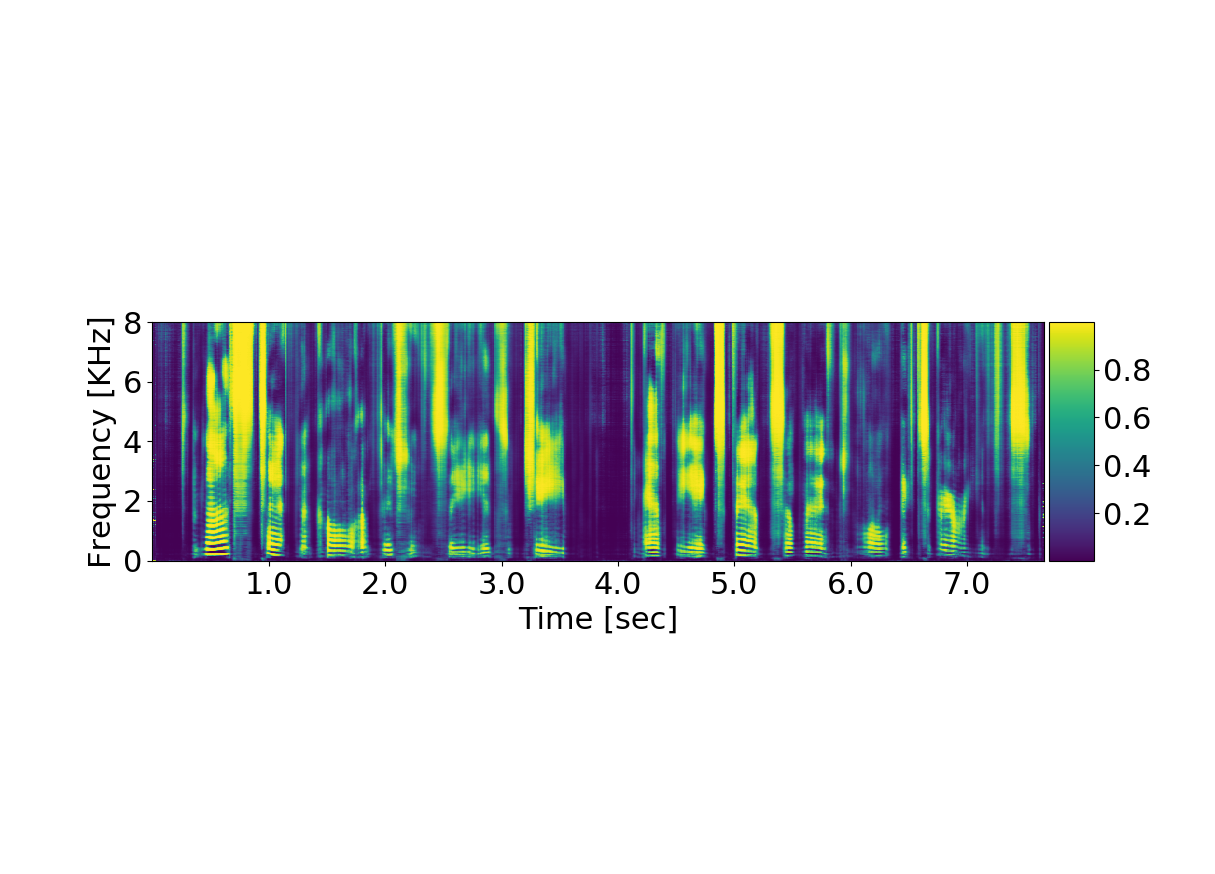}
			\caption{ DMoE-2 SPP.}
			\label{fig:spp_dmoe}
		\end{subfigure}%
		\caption{SPP estimations of Factory noise with SNR = 5dB.}
		\label{fig:factory_results}
	\end{figure*}

	In order to gain further insights into the  capabilities of the proposed algorithm we  compared the enhancement performance of the \ac{DMoE} algorithm with the challenging Factory noise environment. We set the SNR in this experiment to $5$~dB. Fig. \ref{fig:factory_results} depicts the  \ac{SPP} comparison between the \ac{DSE}, the S-\ac{DMoE} and the proposed \ac{DMoE} algorithm performance. Clearly the DSE is very noisy, and the DMoE is smoother than the S-DMoE in both the time and frequency domains.

		\begin{figure*}[tbhp]
				\centering
				\begin{subfigure}[b]{0.5\textwidth}
					%				\centering
					\includegraphics[width=\textwidth]{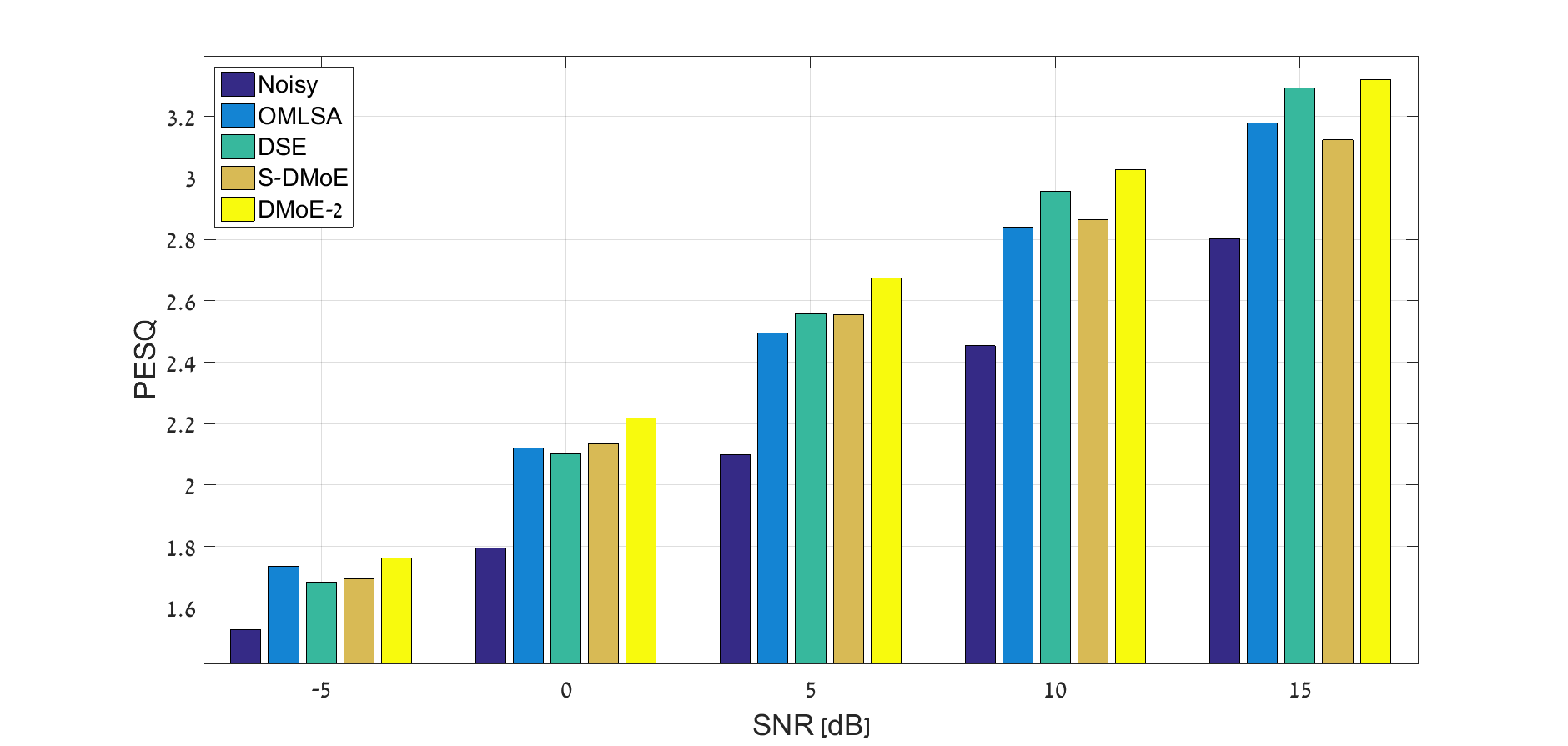}
					\caption{ {Speech} noise.}
					\label{fig:wsj_factory}
				\end{subfigure}%
				\begin{subfigure}[b]{0.5\textwidth}
					%				\centering
					\includegraphics[width=\textwidth]{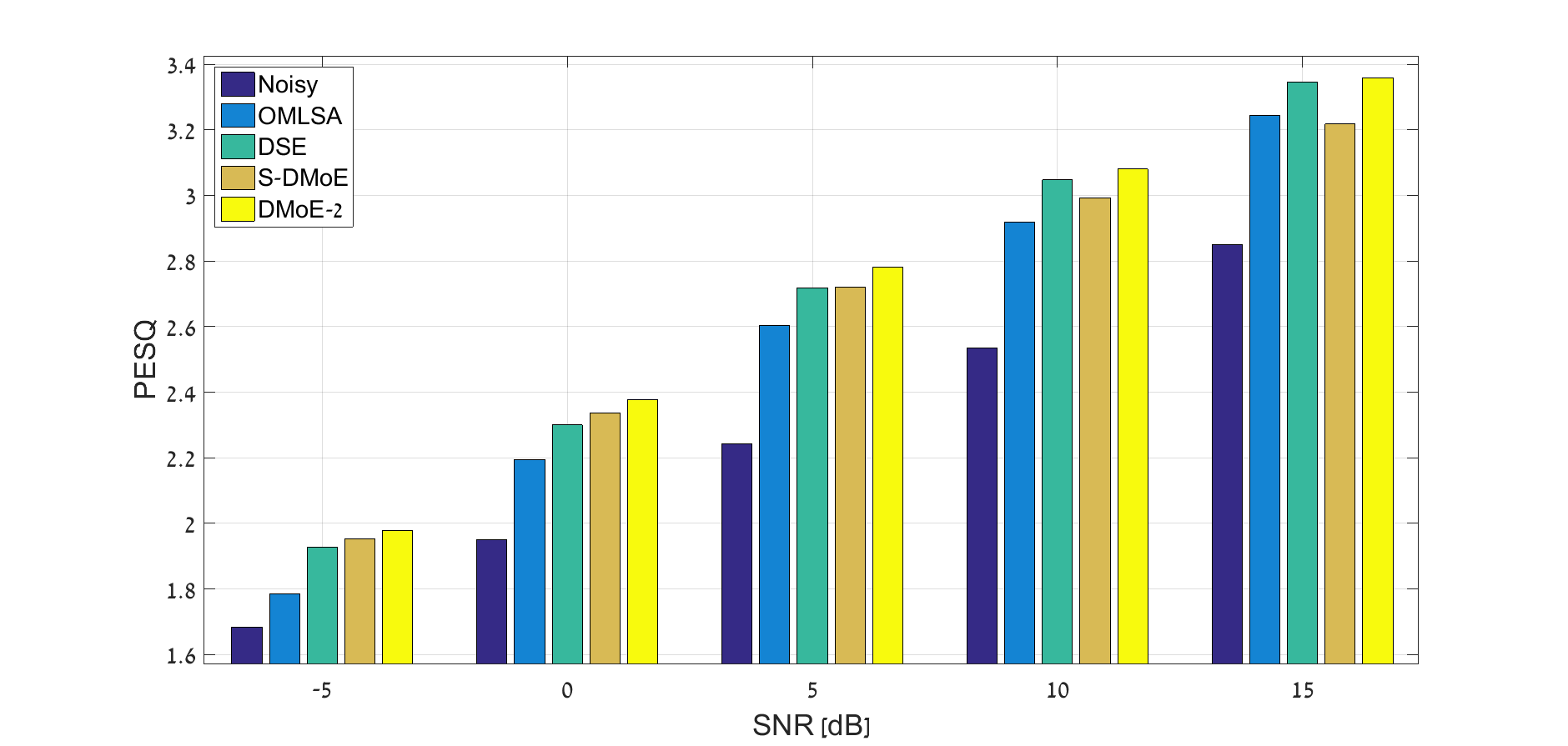}
					\caption{ {Room} noise.}
					\label{fig:wsj_room}
				\end{subfigure}\\		
				\begin{subfigure}[b]{0.5\textwidth}
					\includegraphics[width=\textwidth]{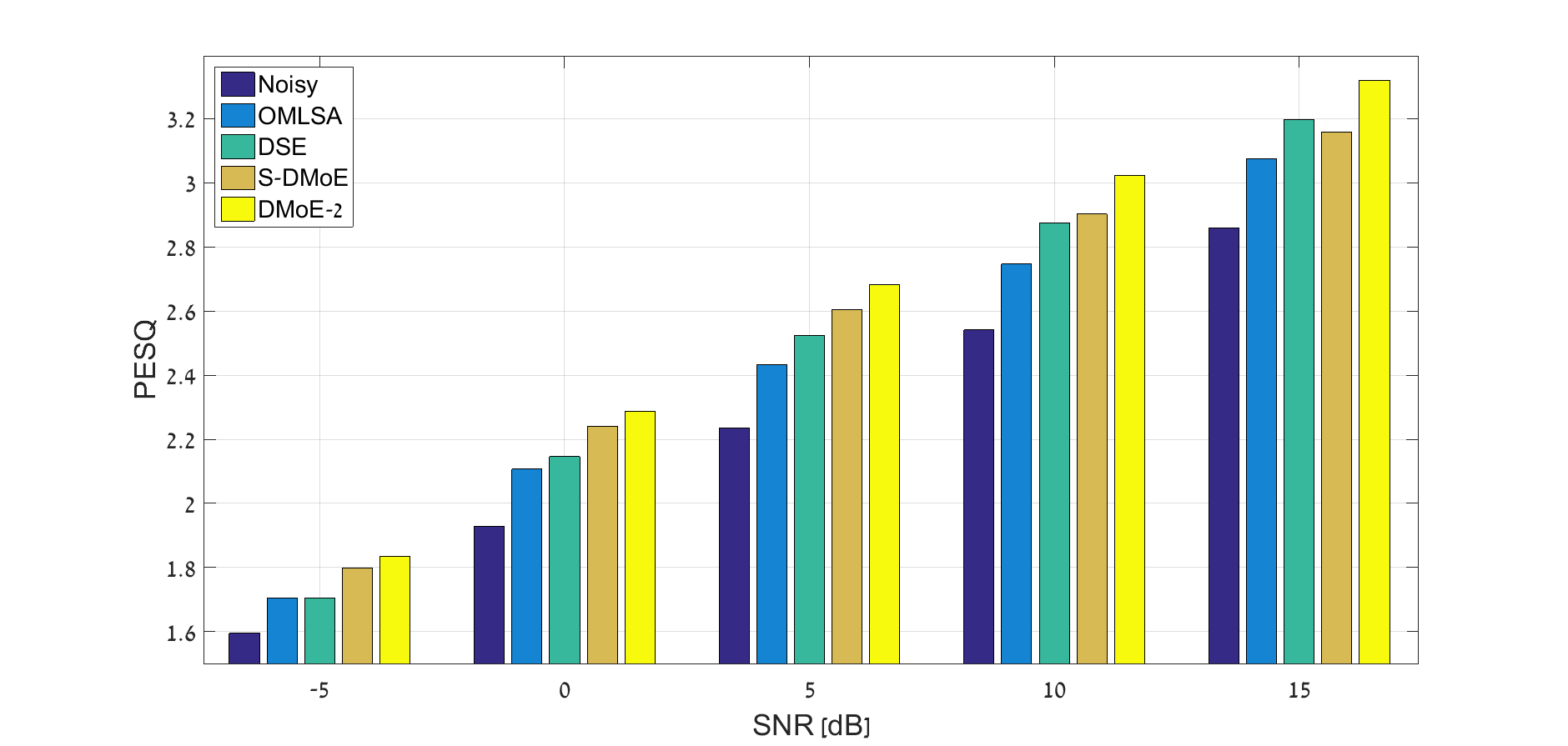}
					\caption{{Factory} noise.}
					\label{fig:wsj_babble}
				\end{subfigure}%
				
				\caption{Speech quality results (PESQ) computed on the \ac{WSJ} database.}
				\label{fig:WSJ}
			\end{figure*}
	\subsection{Performance with a different database}	\label{subsec:notimit}
		Here we have trained the \ac{DMoE} using the TIMIT database. In order to show that the proposed algorithm is immune from the overfitting phenomena we tested the capabilities of the proposed \ac{DMoE} algorithm when applied to speech signals from other databases. We applied the algorithm to $30$ clean signals drawn from the \ac{WSJ} database~\cite{22}. The signals were contaminated by Speech, Room and Factory noises, drawn from the NOISEX-92 database,  with several \ac{SNR} levels. Note, that the algorithm was neither trained on that database nor trained with these noise types. Fig.~\ref{fig:WSJ} depicts the \ac{PESQ} measure of the \ac{DMoE} algorithm in comparison to the other  algorithms. It is evident that the performance of proposed algorithm was maintained even for sentences drawn from a database other than the training database. The results for other noise types,  not shown here due to space constraints, were comparable.

	\section{Discussion}\label{sec:disccusion}
	In this section, we first discuss the role of the number of experts in Sec~\ref{subsec:num_of_experts}. The experts' performance is tested in Sec~\ref{subsec:experts}, and finally, the gating is analyzed in Sec~\ref{subsec:gating}.
	
	\subsection{Setting the number of experts}\label{subsec:num_of_experts}
\begin{figure*}[tbhp]
		\centering
		\begin{subfigure}[b]{0.5\textwidth}
			%				\centering
			\includegraphics[width=\textwidth]{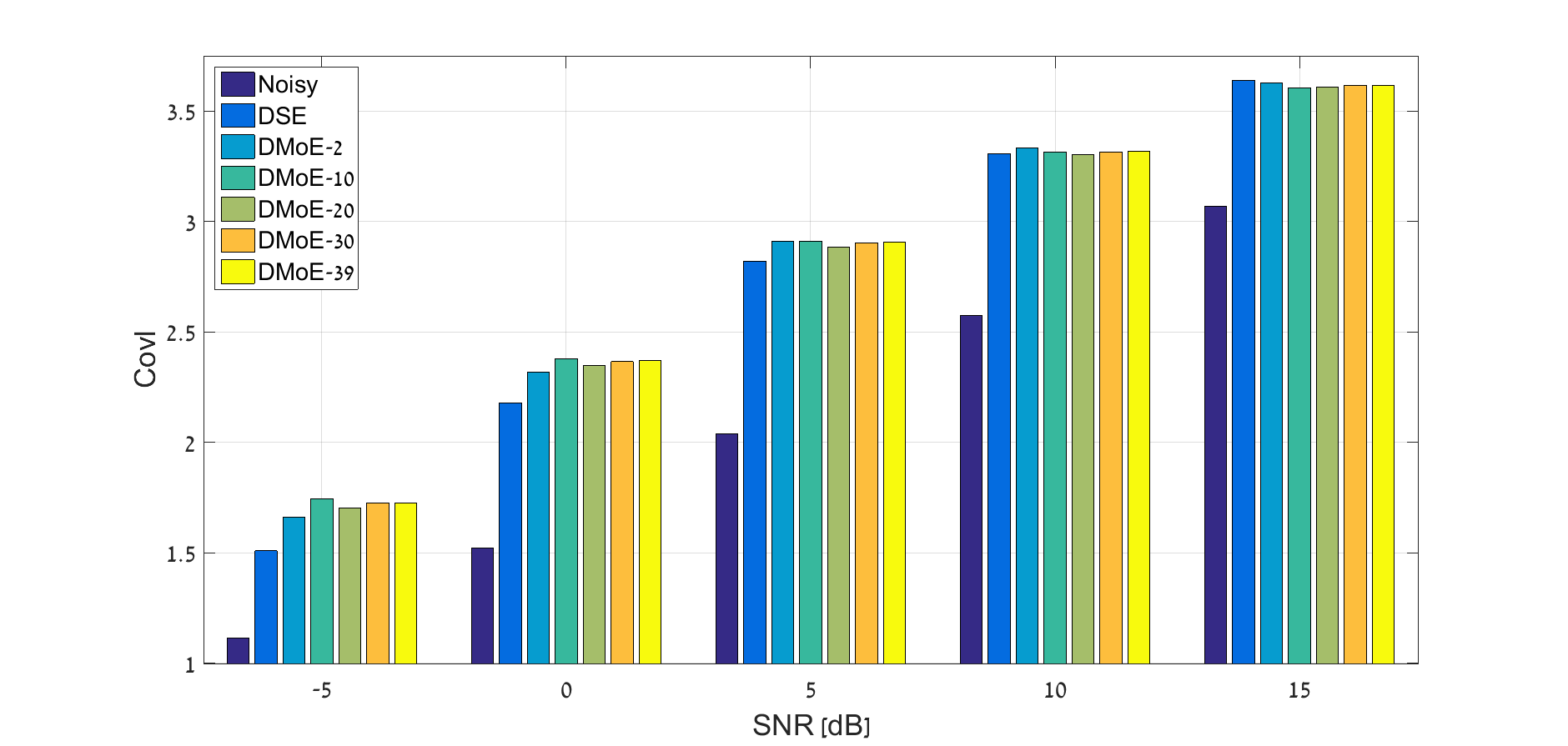}
			\caption{ {Speech} noise.}
			\label{fig:noe_Speech}
		\end{subfigure}%
		\begin{subfigure}[b]{0.5\textwidth}
			%				\centering
			\includegraphics[width=\textwidth]{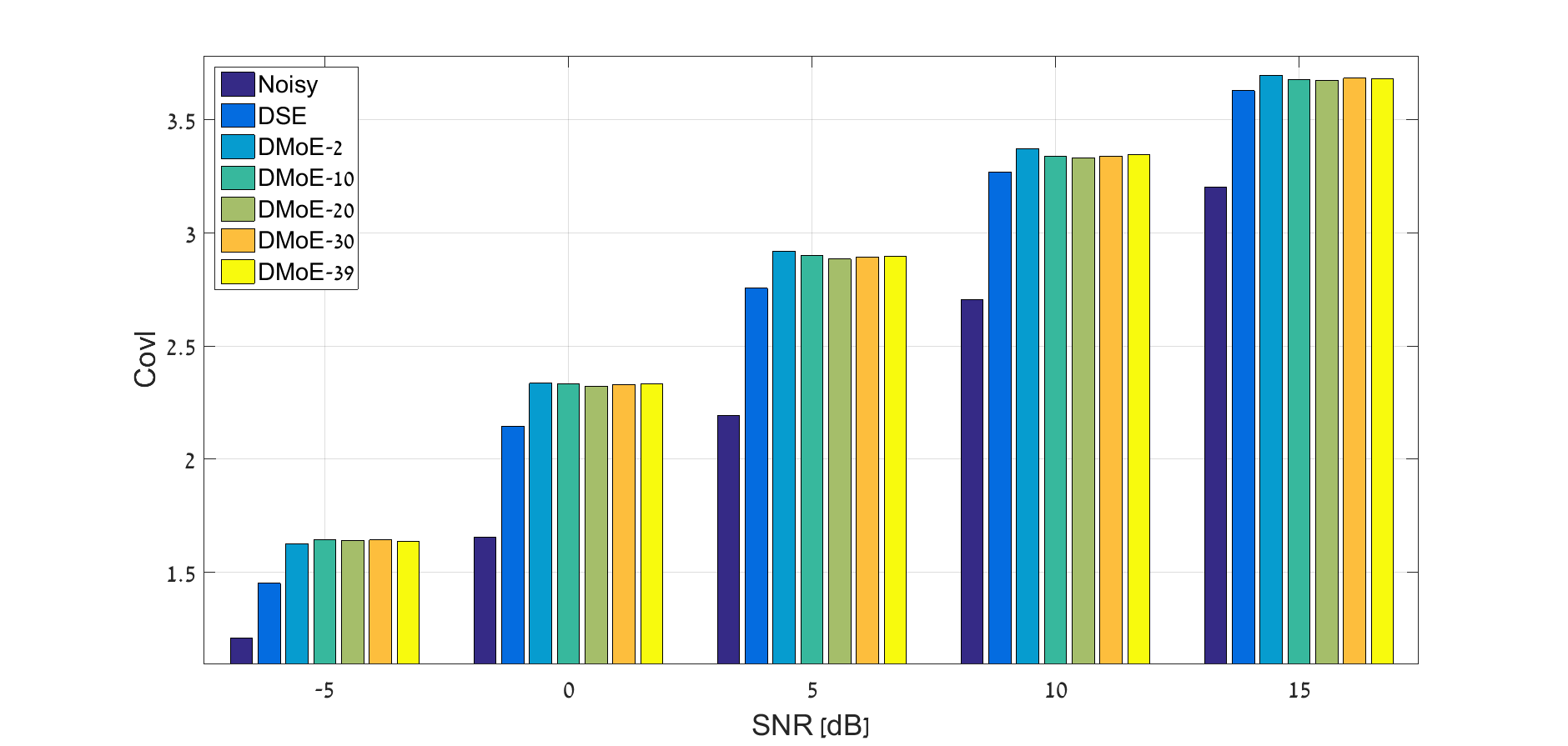}
			\caption{ {Room} noise.}
			\label{fig:noe_Room}
		\end{subfigure}\\		
		\begin{subfigure}[b]{0.5\textwidth}
			\includegraphics[width=\textwidth]{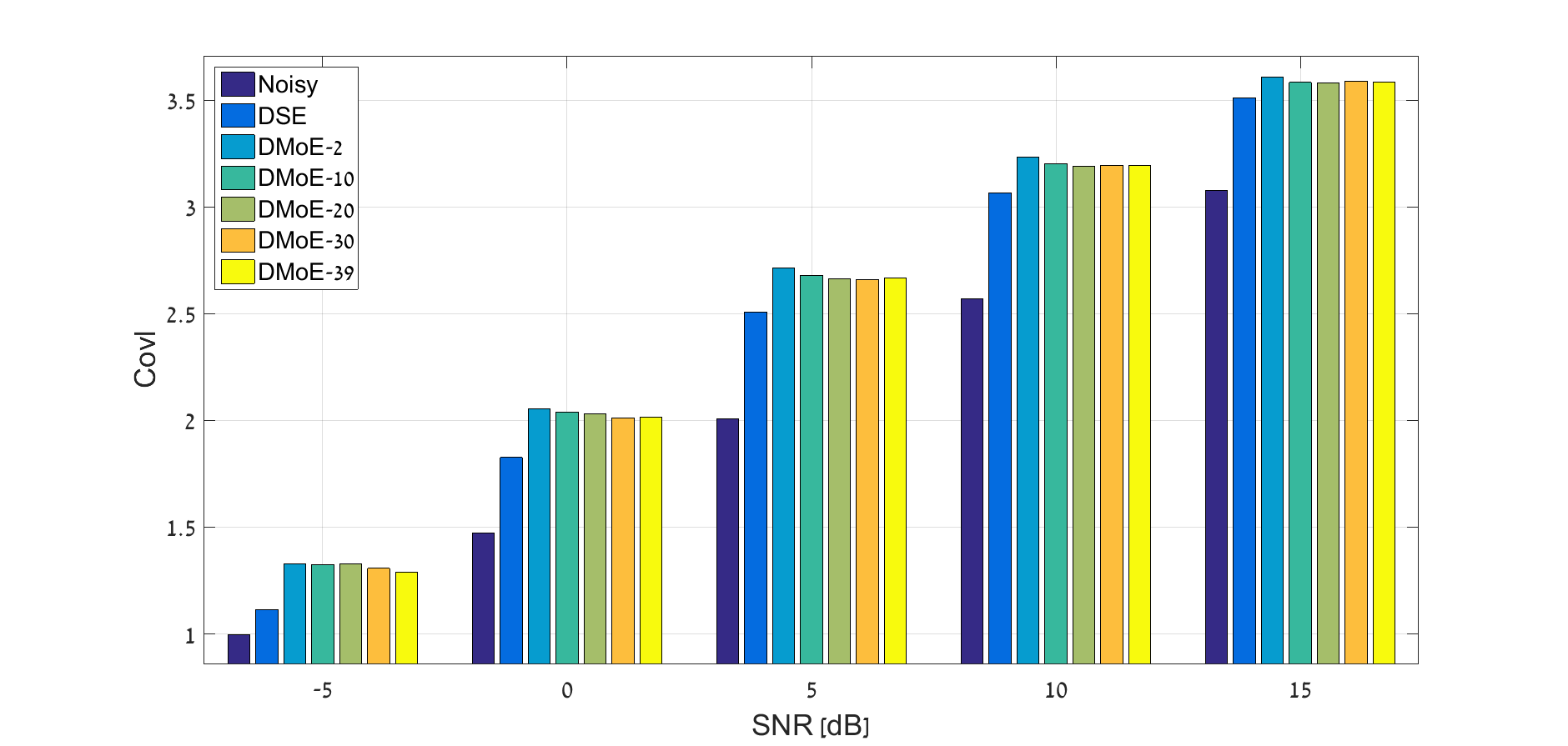}
			\caption{{Factory} noise.}
			\label{fig:noe_factory}
		\end{subfigure}%
		\begin{subfigure}[b]{0.5\textwidth}
			%				\centering
			\includegraphics[width=\textwidth]{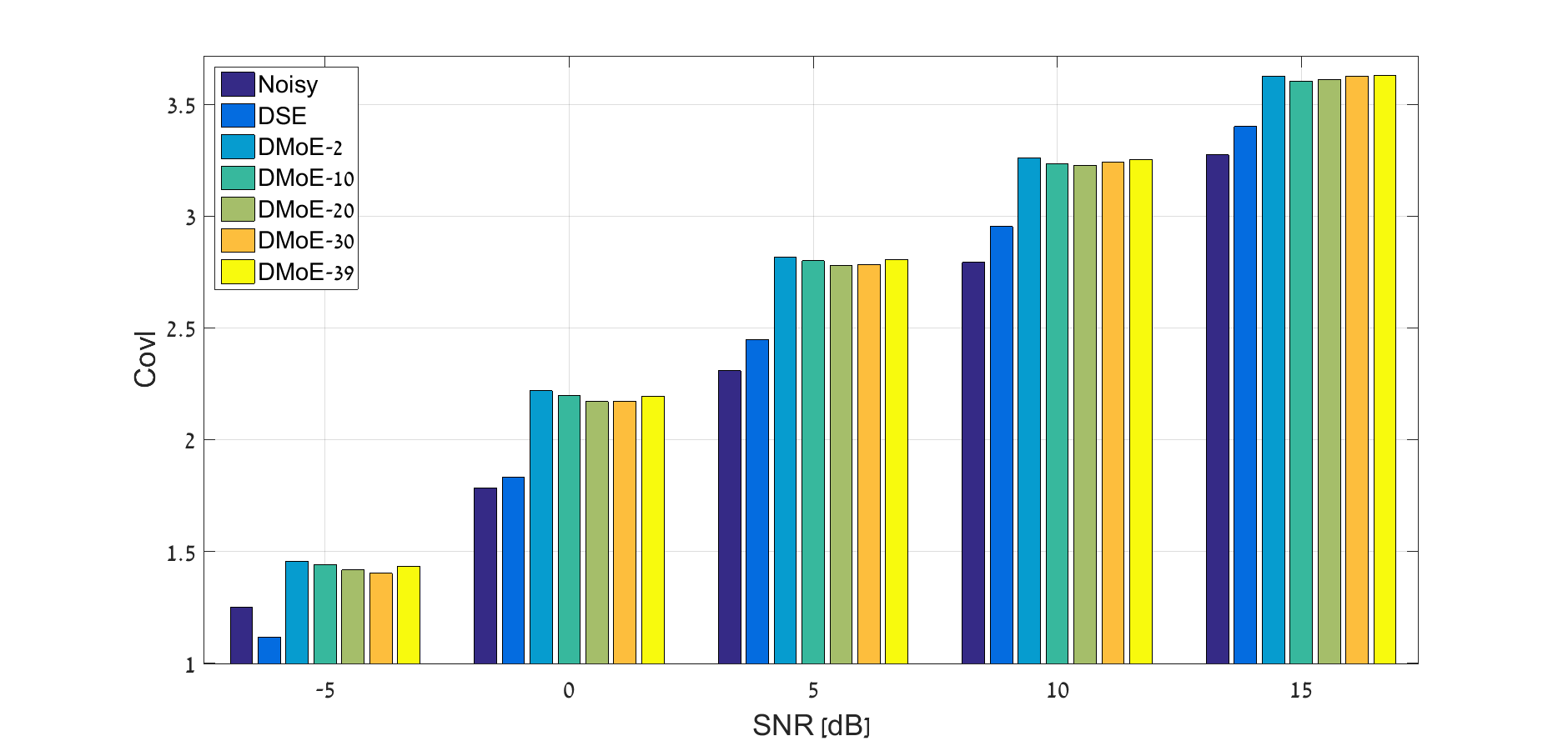}
			\caption{ {Babble} noise.}
			\label{fig:noe_Babble}
		\end{subfigure}%
		\caption{Speech quality results (Covl) for several noise types as a function of the number of experts.}
		\label{fig:noe_Covl}
	\end{figure*}

\begin{figure*}[tbhp]
		\centering
		\begin{subfigure}[b]{0.5\textwidth}
			%				\centering
			\includegraphics[width=\textwidth]{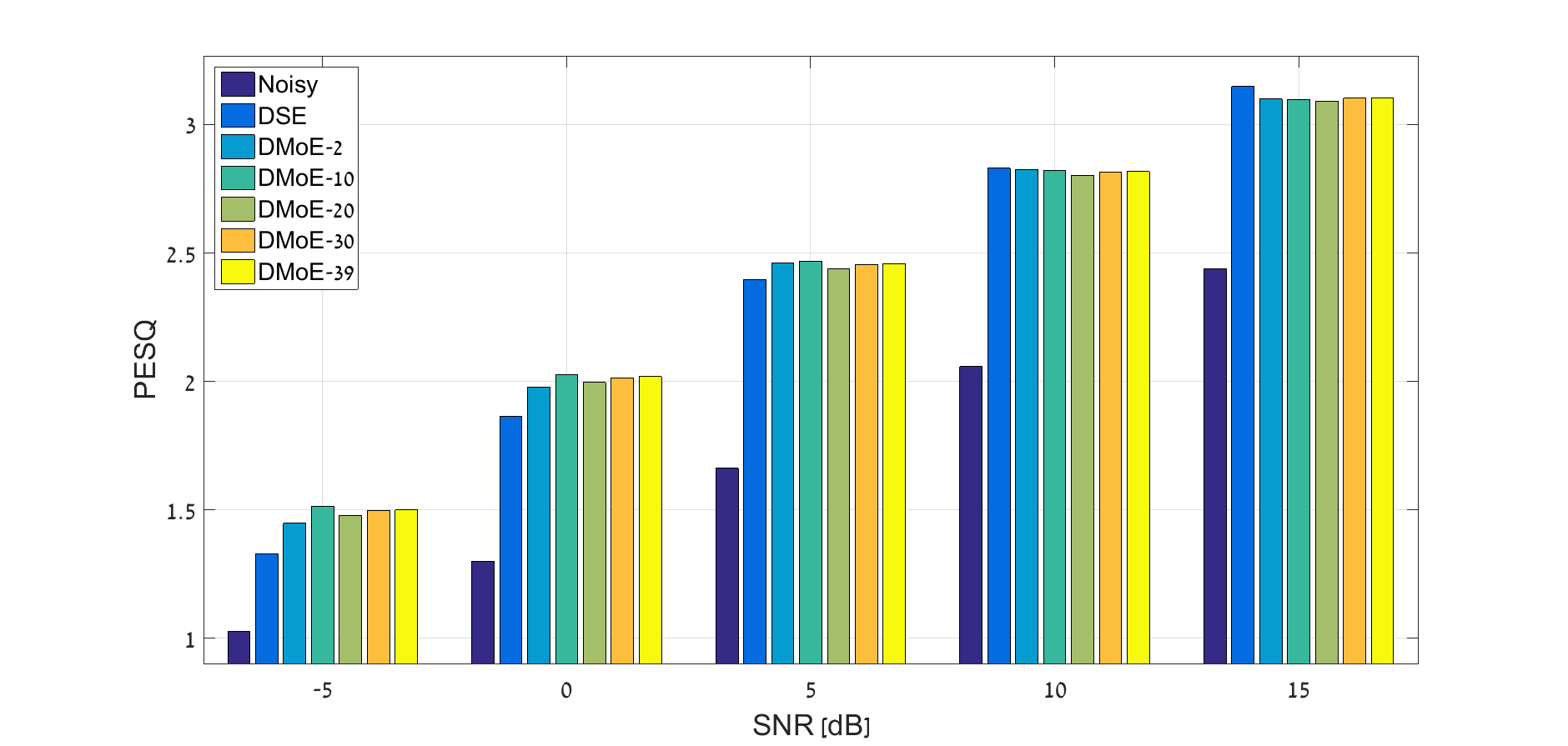}
			\caption{ {Speech} noise.}
			\label{fig:noe_pesq_Speech}
		\end{subfigure}%
		\begin{subfigure}[b]{0.5\textwidth}
			%				\centering
			\includegraphics[width=\textwidth]{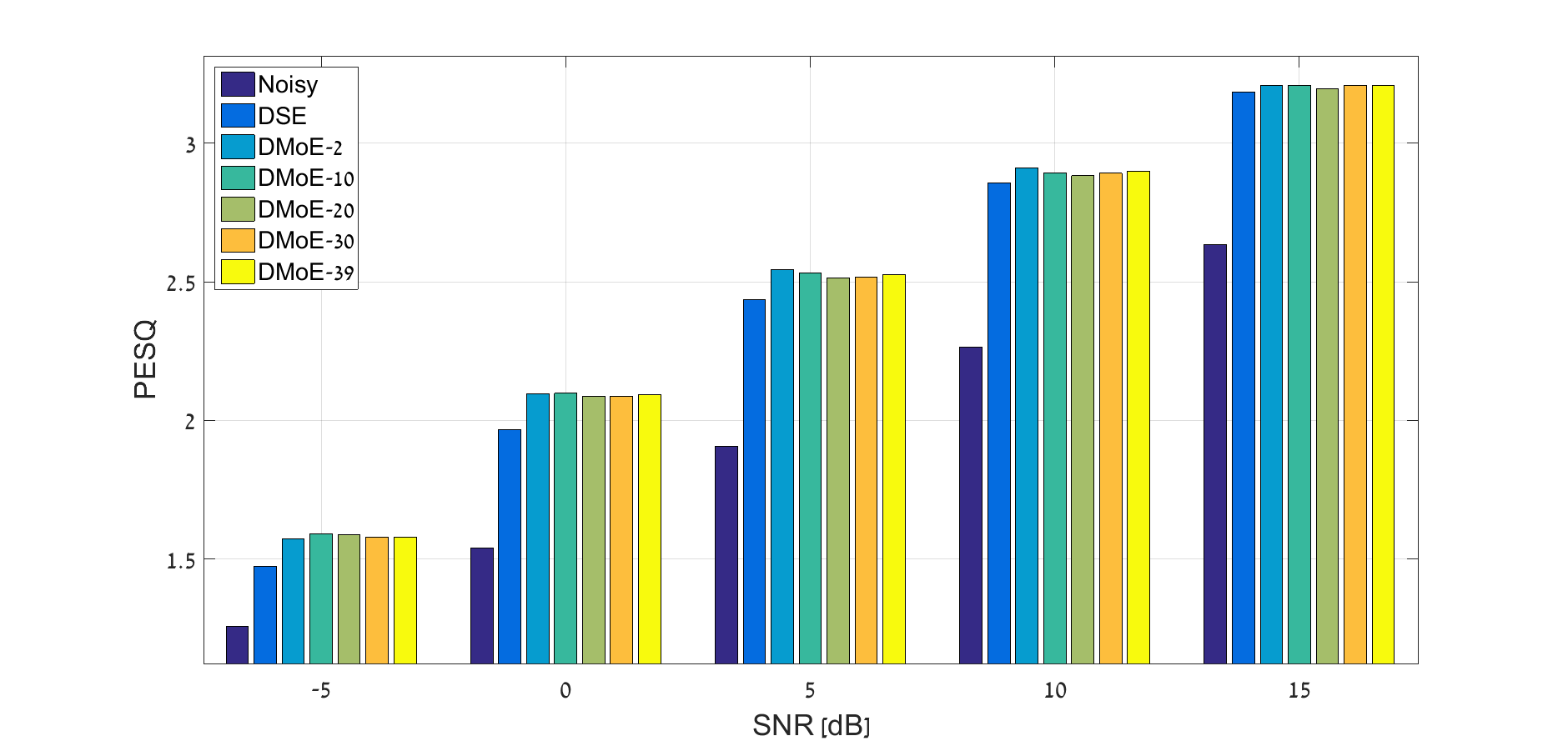}
			\caption{ {Room} noise.}
			\label{fig:noe_pesq_Room}
		\end{subfigure}\\		
		\begin{subfigure}[b]{0.5\textwidth}
			\includegraphics[width=\textwidth]{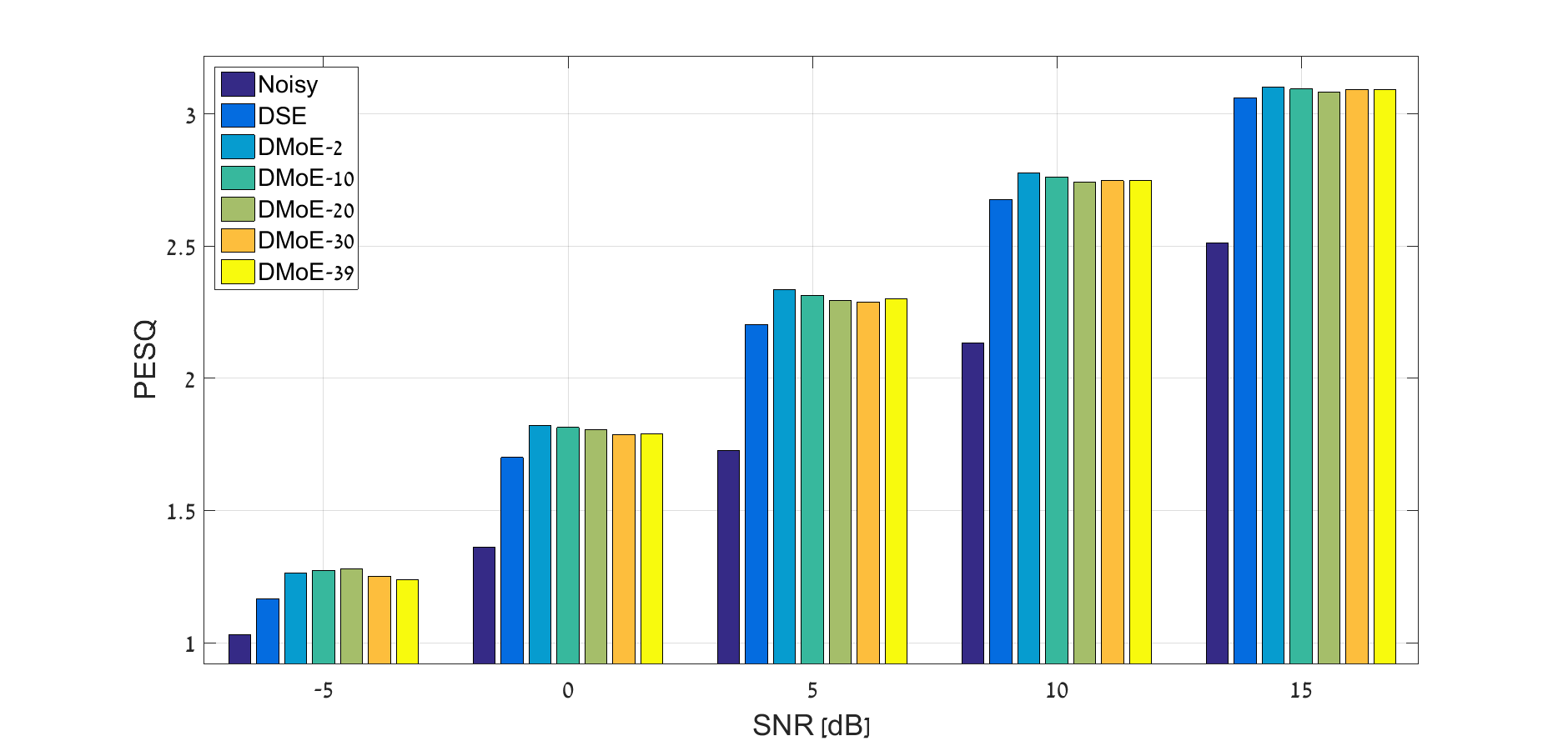}
			\caption{{Factory} noise.}
			\label{fig:noe_pesq_factory}
		\end{subfigure}%
		\begin{subfigure}[b]{0.5\textwidth}
			%				\centering
			\includegraphics[width=\textwidth]{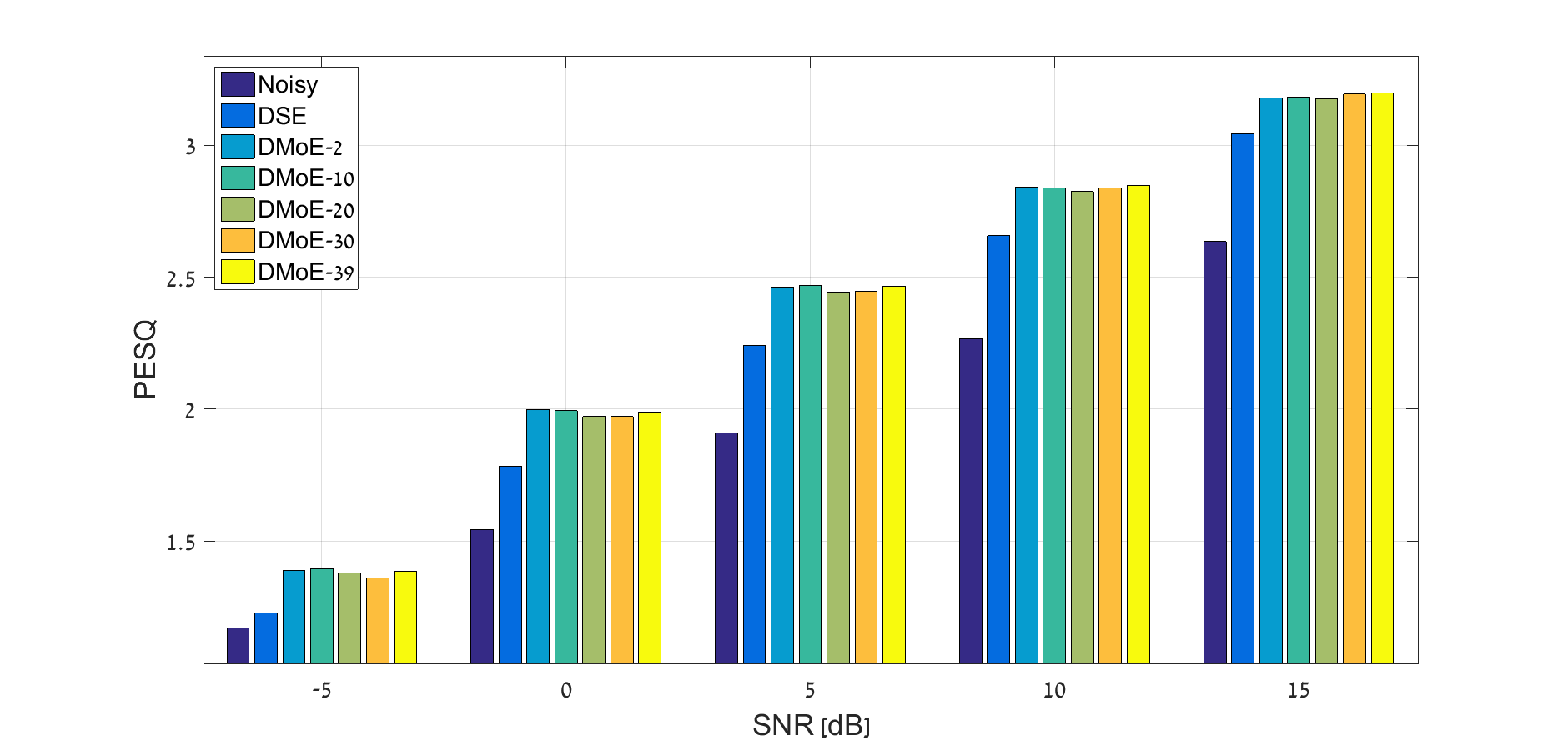}
			\caption{ {Babble} noise.}
			\label{fig:noe_pesq_Babble}
		\end{subfigure}%
		\caption{Speech quality results (PESQ) for several noise types as a function of the number of experts.}
		\label{fig:noe_PESQ}
	\end{figure*}

	The  proposed architecture is based on a mixture of $m$ experts and the gating network directs a given input to one of these $m$ experts.
In most  MoE studies, finding the number of experts   was done by exhaustive search \cite{yuksel2012twenty}.
In our case we divide the (log-spectrum) feature space into  $m$ simpler subspaces.
    On one hand, setting a large value for $m$  divides the problem into many experts thus giving each expert an easier job of enhancing a distinct speech type. On the other hand, when $m$ is large the model complexity is higher, which makes the training task more difficult. Additionally, when the model size is large, the computational demands can make it difficult to use  in real time applications.
	In order to determine the best value of $m$ based on enhancement performance, we conducted an experiment using   \ac{DMoE} models   with  1, 2, 10, 20, 30 and 39 experts. We use the notation  \ac{DMoE}-$m$ for a \ac{DMoE} based on $m$ experts. The 6 architectures were trained separately on the train part of the TIMIT database. We  then tested the obtained networks using the same procedure described in Sec~\ref{subsec:expsetup}.
	Fig. \ref{fig:noe_Covl} shows the \ac{Covl} results, and Fig. \ref{fig:noe_PESQ} shows the \ac{PESQ} results. It is clear that moving from a single expert to two experts significantly improves the results. However, as can be seen from  the results, further  increasing the number of experts did not improve the performance much.
	
	\begin{figure*}[tbhp]
				\begin{subfigure}[b]{0.5\textwidth}
			%				\centering
			\includegraphics[trim=0 0 0 0 ,clip,width=\textwidth]{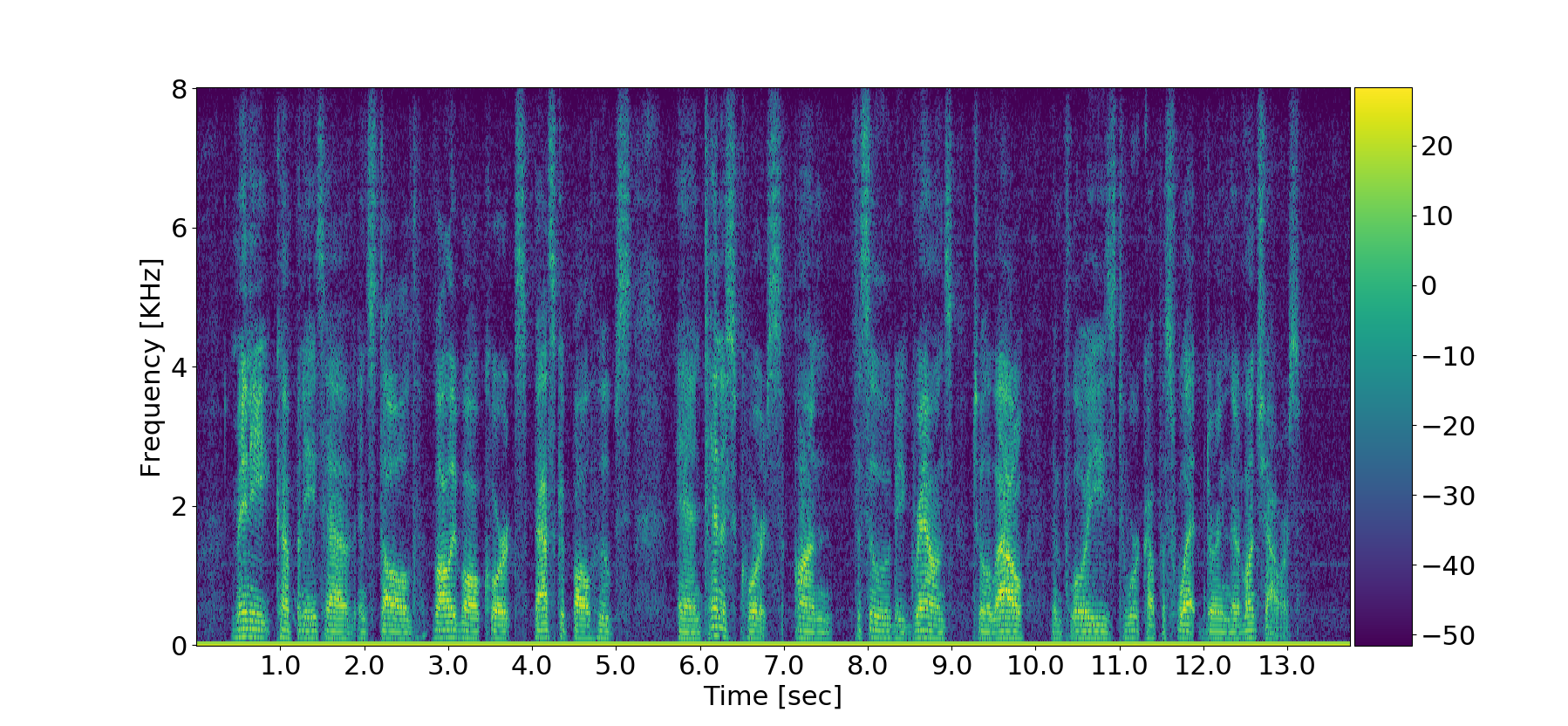}
			\caption{Clean speech.}
			\label{fig:exp_clean}
		\end{subfigure}
	\begin{subfigure}[b]{0.5\textwidth}
			%				\centering
			\includegraphics[trim=0 0 0 0 ,clip, width=\textwidth]{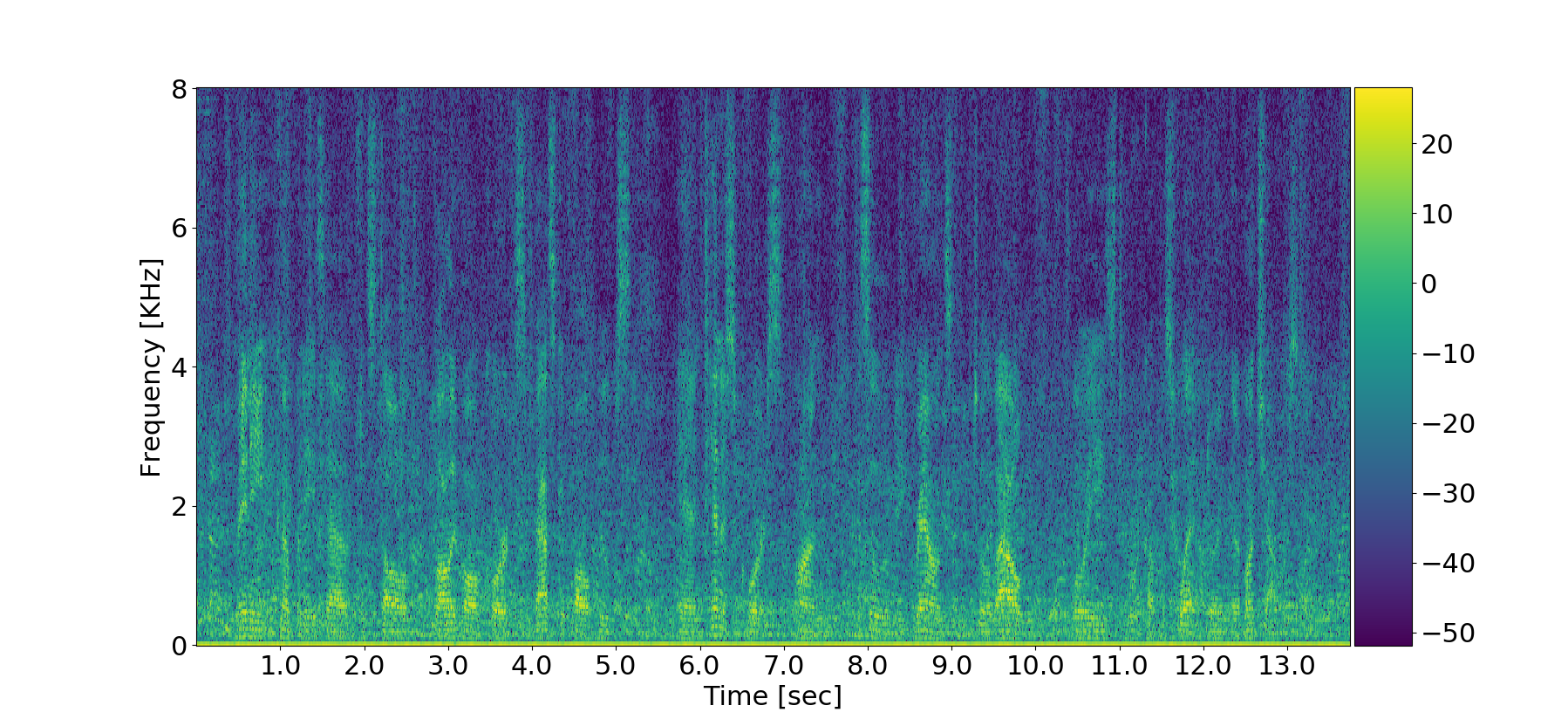}
			\caption{ Noisy speech.}
			\label{fig:exp_noisy}
		\end{subfigure}\\
		\begin{subfigure}[b]{0.5\textwidth}
			%				\centering
			\includegraphics[trim=0 0 0 0 ,clip, width=\textwidth]{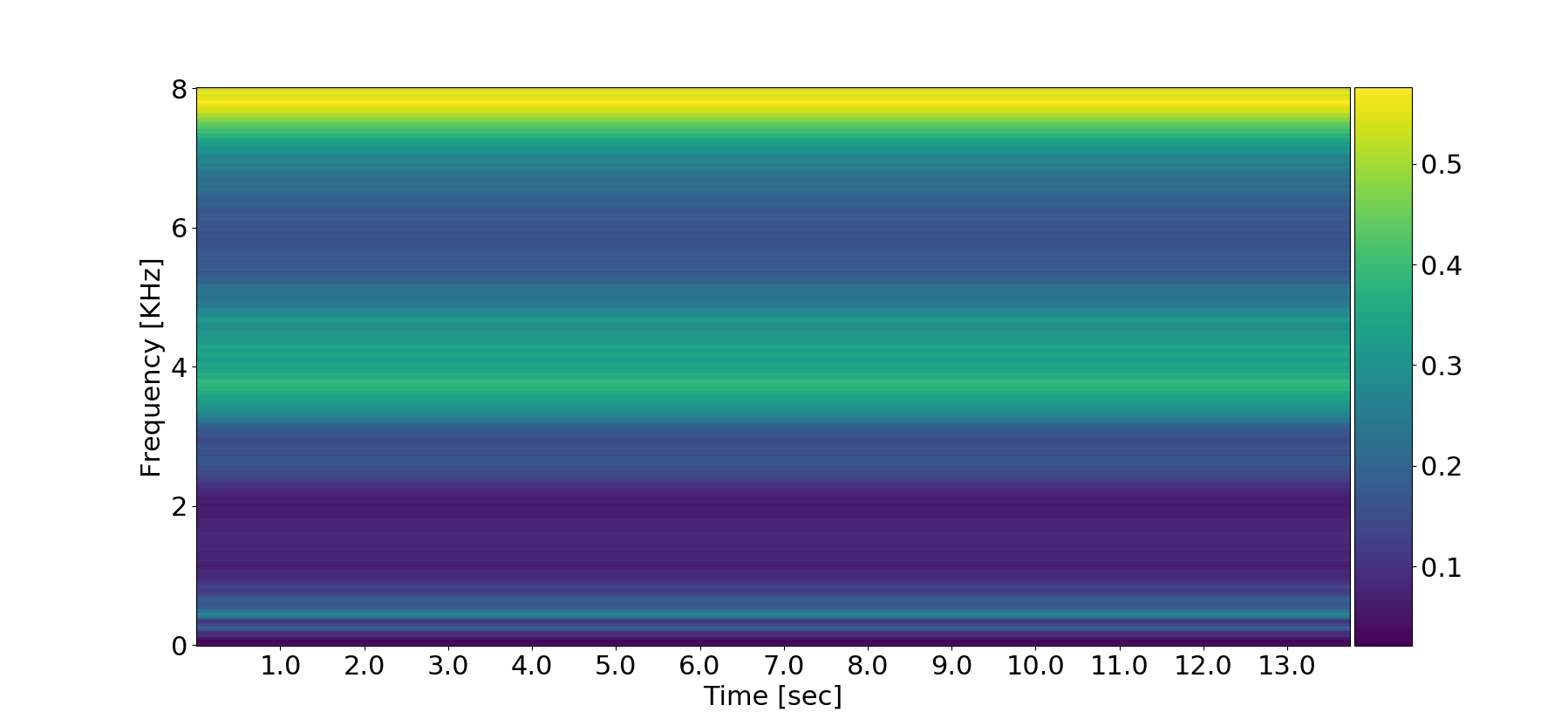}
			\caption{ DSE SPP.}
			\label{fig:exp_single}
		\end{subfigure}%
		\begin{subfigure}[b]{0.5\textwidth}
			%				\centering
			\includegraphics[trim=0 0 0 0 ,clip,width=\textwidth]{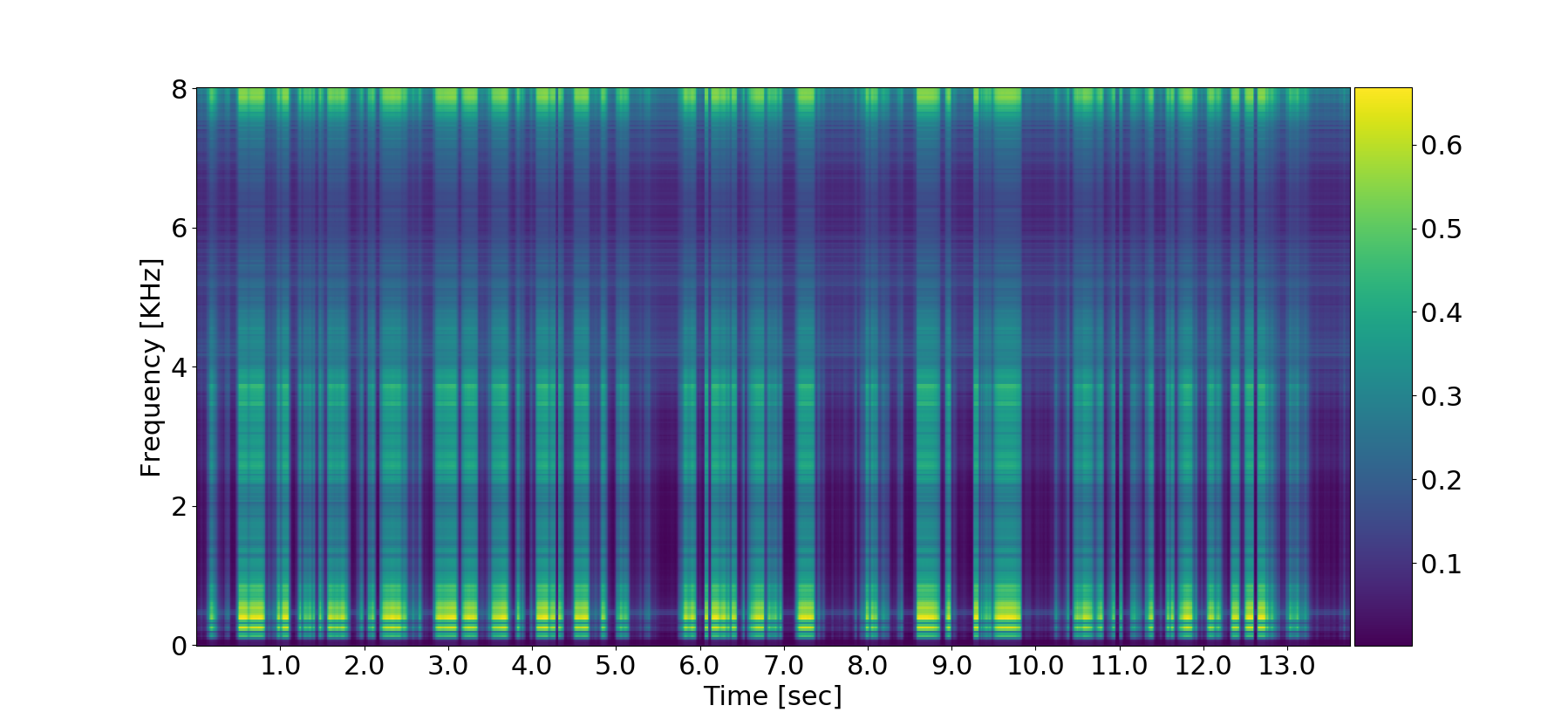}
			\caption{DMoE-2 SPP.}
			\label{fig:exp_2e}
		\end{subfigure}	\\
		 \begin{subfigure}[b]{0.5\textwidth}
			\centering 	\includegraphics[trim=0 0 0 0 ,clip,width=\textwidth]{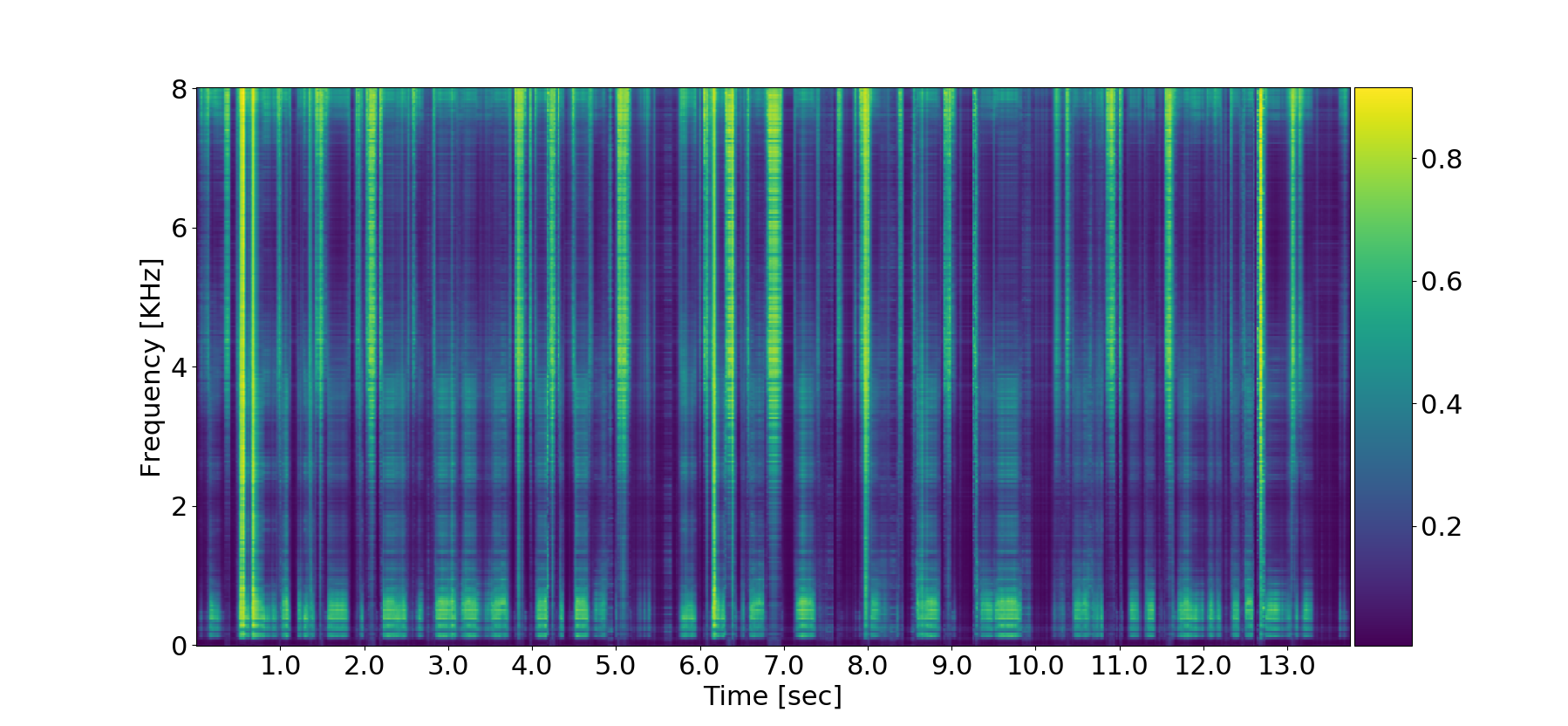}
			\caption{DMoE-30 SPP.}
			\label{fig:exp_30e}
		\end{subfigure}
		\caption{Enhancement performance based solely on the input data to the gating network (and using a fixed non-informative vector as  input to the experts' networks). We show the estimated SPP for several speech enhancement networks. }
		\label{fig:experts}
	\end{figure*}

	\subsection{The experts' expertise}\label{subsec:experts}
	To  illustrate the role of different  experts,  we added non-stationary Factory noise to clean speech from the TIMIT database with SNR=5 dB. The gating \ac{DNN} was fed with the \ac{MFCC} features of the noisy input and produced the probability of each  expert. Now, instead of introducing the experts to the noisy log-spectrum vector $\x$, a vector of all-ones was used as  input. This non-informative vector  can be viewed as a noise-only signal. This vector propagated through the experts and in the end, the SPP $\rrho$ was estimated.
	Fig. \ref{fig:experts}  depicts the output of the enhancing procedure.
Figs. \ref{fig:exp_clean} and \ref{fig:exp_noisy} show the clean and noisy speech.
Fig. \ref{fig:exp_single} shows the estimated \ac{SPP} of the fully-connected (single expert) architecture. It is clear that since the input is the same and the \ac{MFCC} gating-decision data are irrelevant here, no information is preserved and the results therefore are meaningless. Fig. \ref{fig:exp_2e} shows the estimated \ac{SPP} of the \ac{DMoE}-2 architecture. Here, the extra information of the gating helps the experts to follow the structure of the input. Note, that only two different structures are shown here, based on  the two experts. Simple inspection shows  that one SPP structure appears in the voiced frames and the other in the unvoiced frames. Finally, in Fig. \ref{fig:exp_30e} the estimated \ac{SPP} from \ac{DMoE}-30 is depicted. It is easy to see that this architecture tracks the clean speech more accurately. Now that more experts are present,  more speech structures are estimated.
	
	This experiment suggests that each expert is responsible for a specific structure of the speech. Consequentially, the experts preserve the speech structure even if introduced to an unfamiliar noise. This leads to  more robust behavior compared to other \ac{DNN}-based algorithms.

	\subsection{The gating voice/unvoiced decision}\label{subsec:gating}
\begin{figure}[tbhp]
		\centering
		\begin{subfigure}[b]{0.5\textwidth}
			%				\centering
			\includegraphics[trim=0 220 0 220 ,clip, width=\textwidth]{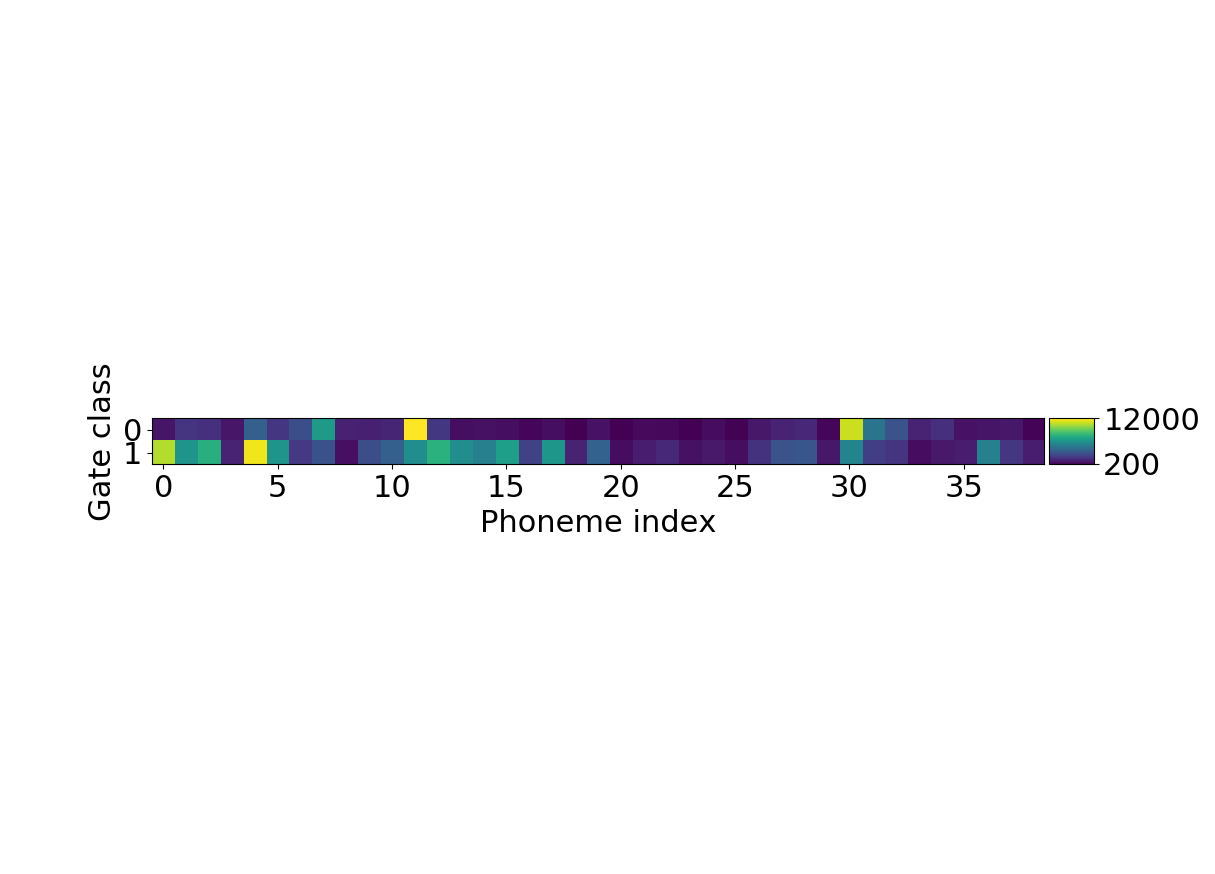}
			\caption{ The binary gating  distribution  for each  phoneme.}
			\label{fig:conf}
		\end{subfigure}\\%
		\begin{subfigure}[b]{0.5\textwidth}
			%				\centering
			\includegraphics[trim=0 0 0 35 ,clip,width=\textwidth]{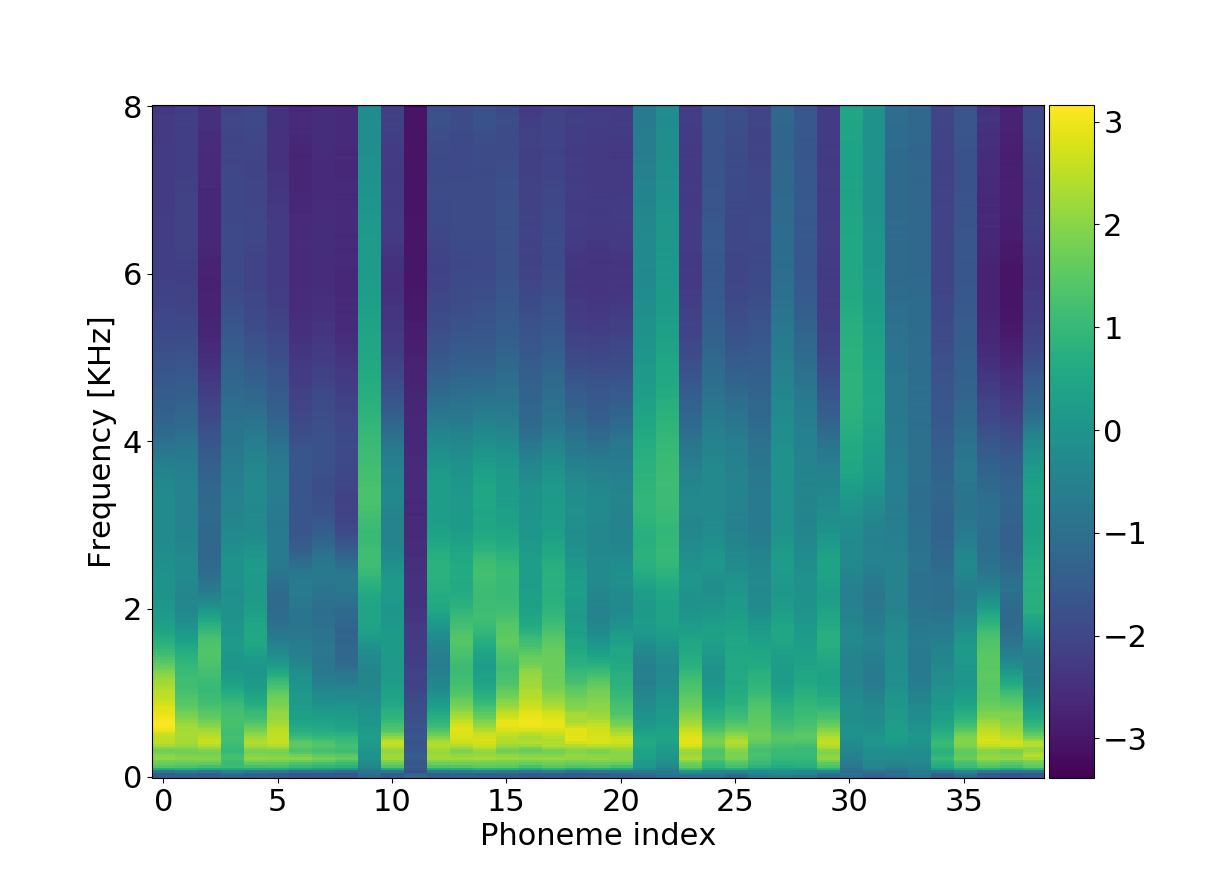}
			\caption{ Phonemes \ac{PSD}.}
			\label{fig:mus}
		\end{subfigure}		
		\caption{Average phonemes  distribution in the log-spectrum domain.}
		\label{fig:conf &mus}
	\end{figure}

 Fig.  \ref{fig:exp_2e} depicts the case of two experts, one  responsible for  enhancing  voiced frames and the other for the unvoiced frames.
The gating network therefore needs to classify the frame as  voiced or unvoiced in order to direct it to the appropriate expert. We next show in a more systematic and quantitative manner that this is indeed what the gate network does.	Note, that in the training phase no phoneme labels are provided to the gating \ac{DNN}. The TIMIT database is  phoneme labeled. Hence we can collect  statistics on gate decisions for each phoneme separately. 	
	Fig. \ref{fig:conf} depicts the gate decision statistic as a function of the phoneme label, and  Fig. \ref{fig:mus} presents the average structure of the phonemes in the log-spectrum domain. It appears from the gating decisions  that the gating \ac{DNN} tends to direct unvoiced phonemes to one expert and the voiced phonemes   to the second. This partition, which is obtained in an unsupervised manner, makes sense since the voice/unvoiced structures are dramatically different:  the voiced phonemes are characterized by energy in the low frequencies, whereas the unvoiced phonemes are characterized by energy in the high frequencies.

%	\subsection{small database and noises}
%	Usually, when training a \ac{DNN} big database is required, especially with complex problems. The massive amount of data is used in order to  minimize the mismatch between the training and the test conditions. In \cite{wang2015large} for example,  massive  of noise time were used to train a fully-connected \ac{DNN} for speech enhancement task. The training of-course is very long. Furthermore when an unfamiliar noise background appears  the fully-connected \ac{DNN} performance desegregates.
%	
%	In our approach, since each expert is trained on easier task the variability of the input is lessened and therefore even with an unfamiliar noise the experts will succeed.
	
%	\subsection{Speech structure is preserved}
%	
%	\subsection{robustness to different }
%	
%	\subsection{reconstruction of vanished information due to high SNRs }

	\section{Conclusion} \label{sec:summery}
	This article introduced a \ac{DMoE} model for speech enhancement. This approach divides the  challenging task of speech enhancement into simpler tasks where each \ac{DNN} expert is responsible for a  simpler one. The gating \ac{DNN} directs the input features to the correct expert.
We showed empirically that in the case of two experts, the gating decision is correlated with  the voice/unvoiced status of the input frame.

	In a fully connected \ac{DNN}, the input to a single \ac{DNN} has to deal with both  voiced and unvoiced frames, which leads to  performance degradation. On the other hand, in  the \ac{DMoE} model, the gating splits the problem into simpler problems and each expert preserves the pattern of the the spectral structure of frames that are directed to this expert. 
		This approach makes is possible to  overcome the well known  problem of \ac{DNN}-based algorithms; namely, the  mismatch between training phase and test phase. Additionally, the proposed \ac{DMoE} architecture enables training with a small database of noises.
	
	The experiments showed that the proposed algorithm outperforms state of the art algorithms as well as \ac{DNN}-based approaches on the basis of objective and subjective measurements.

	\balance
	
	\bibliographystyle{IEEEtran}
	\bibliography{main}

\end{document}